\documentclass[twocolumn]{aastex62}
\bibpunct{(}{)}{;}{a}{}{,}

\shorttitle{XLF of ULXs in RiGs}
\shortauthors{Wolter et al.}

\received{March, 7th, 2018}
\accepted{June, 5th, 2018}
\submitjournal{The Astrophysical Journal}

\begin{document}


\title{The X-Ray Luminosity Function of Ultra Luminous X-Ray Sources in Collisional Ring Galaxies}

\author{Anna Wolter}
\affil{INAF-Osservatorio Astronomico di Brera, via Brera 28, 20121 Milano, Italy}
\author{Antonella Fruscione} 
\affil{Harvard-Smithsonian Center for Astrophysics, 60 Garden St., Cambdidge, MA, 02138}
\author{Michela Mapelli}
\affil{INAF-Osservatorio Astronomico di Padova, vicolo dell'Osservatorio 5, 35122, Padova, Italy }







\begin{abstract}
Ring galaxies are fascinating laboratories: a catastrophic impact between two galaxies (one not much smaller than the other) has produced fireworks especially in the larger one, when hit roughly perpendicularly to the plane. 
We analyze the point sources, produced by the starburst episode following the impact, in the rings of seven galaxies and determine their X-ray luminosity function (XLF). In total we detect 63 sources, of which 50 have luminosity L$_X \geq 10^{39}$ erg s$^{-1}$, classifying them as ultra luminous X-ray sources (ULXs). We find that the total XLF is not significantly different from XLFs derived for other kinds of galaxies, with a tendency of having a larger fraction of high X-ray luminosity objects. Both the total number of ULXs and the number of ULXs per unit star formation rate are found in the upper envelope of the more normal galaxies distribution. Further analysis would be needed to address the issue of the nature of the compact component in the binary system.
\end{abstract}





\section{Introduction}
\label{intro}

Ring Galaxies (RiGs hereafter) are unique laboratories in which to study peculiar phases of galaxy evolution. At least a fraction of ring galaxies are thought to form via (almost) head-on collisions with massive intruder galaxies. Due to the gravitational perturbation induced by the bullet galaxy, a density wave propagates through the disc of the target galaxy, generating an expanding ring of gas and stars \citep[e.g.][]{Lynds76}.
They are often characterized by high star formation rates
(SFR$\approx{}0.1-20$ M$_\odot{}$ yr$^{-1}$), suggesting that the
density wave associated with the propagating ring triggers the
formation of stars. Even if they constitute a small subclass of 
galaxies (they are estimated to be about 0.02-0.2 percent of all spiral galaxies, \citealt{Athanassoula85}) 
they represent a peculiar and energetic environment worth investigating.

The archetype of RiGs is the Cartwheel galaxy (see e.g. \citealt{Wolter99,WT04} for a description of the X-ray emission in the Cartwheel and references for multi-wavelength observations). 
It is bright in all bands, and contains a large number of ultra luminous X-ray sources (ULXs), most of which trace the star forming regions in the ring. 

ULXs are off-center point-like sources with X-ray luminosity (assumed isotropic) $L_X>10^{39}$ erg s$^{-1}$, higher than the Eddington limit for a few M$_\odot{}$ black hole (BH).
The general consensus until recently was that most ULXs represent the high luminosity tail of high mass X-ray binaries (HMXBs, see for instance the X-ray luminosity functions of ULX in the Cartwheel, Antennae and NGC 2276 in \citealt{WT04,Zezas07,Wolter15}) 
and the BHs powering them are of (heavy) stellar mass size \citep{Feng11,kaaret17}. 
However, it has not yet been excluded that a number of ULXs are powered by intermediate-mass black holes (IMBH). IMBHs are compact objects with masses in the range $10^2-10^{5}\,{} M_\odot$, which are intermediate between stellar mass BHs and super-massive BHs in the center of galaxies (see e.g. \citet{vanderMarel04}, 
for a review), and have fundamental cosmological implications, as they are deemed to be seeds of super-massive BHs, sources of pre-heating of the intergalactic medium  (e.g. \citealt{Fragos13, Mesinger13}), of fluctuations in the near-infrared
cosmic background \citep{Yue13} and of gravitational waves (e.g. \citealt{Mapelli16,Abbott17c}).
Furthermore, recent observations have found that at least 4 ULXs, including one of the most luminous, are powered by a neutron star (NS), betraying its presence through X-ray pulsations - hence the name PULXs, from pulsar ULXs \citep{Bachetti14, Israel17a, Israel17b, Fuerst16,Carpano18}. Indeed, it might even turn out that all ULXs are NS!  \citep[e.g.][]{King16}.

It is therefore most likely that the class of ULXs, being defined purely on one single observational parameter (their X-ray luminosity), contains a mixed class of sources. 
Mass measure is a very challenging task for ULXs and only few reliable results are available. 
The dynamical evidence for a small mass for the compact object is scanty: it comes mostly from ULX-1 in M101 \citep{Liu13}, and P13 in NGC 7793 \citep{Motch14}. In fact, source P13 turned out to be a NS \citep{Israel17a,Walton18}.
On the other hand, the most luminous ULXs known, and therefore the most promising IMBH candidates, are M82-X1, ESO 243-49 HLX-1, NGC 2276-S6, and Cartwheel-N10 \citep{portegieszwart04,feng10,Farrell09,mezcua15,Wolter06}.

With the aim of studying ULXs of extreme properties in extreme environments, we decided to better investigate a sample of RiGs in search for ULXs. The objects in our sample are similar to the Cartwheel, suggesting the presence of a catastrophic event. 
If RiGs are "caught in the act" of forming new stars, we might be in a better position for finding the most massive products of star formation. 
Ages of rings are estimated to be of not longer than a few hundred Myrs (see Sect.~\ref{galaxies} for age determinations, when available, and references).
Furthermore, due to the peculiar morphology of ring galaxies, point sources detected in the ring are very likely to be physically associated with the galaxy, reducing the problem of contamination from spurious sources.

We collected therefore all collisional RiGs observed by Chandra, four of them already published, three analyzed in this paper for the first time, with the aim of constructing their global X-ray luminosity function (XLF) to study the population of ULXs on statistical grounds. Section 2 describes the sample and the individual galaxies, Section 3 the XLF and Section 4 our discussion of the findings. We use H$_0=73$ Km s$^{-1}$ Mpc$^{-1}$ to compute X-ray luminosities. 

\section{The Sample}
Collisionally formed rings, with 
enhanced star formation, represent an ideal nursery to collect 
a fair sample of ULXs.  And yet only a few X-ray observations of ring 
galaxies exist. We focus on the objects observed by the Chandra satellite because of the superior spatial resolution of the telescope, that allows for a better identification of the sources and less source confusion.

Chandra observed only 7 RiGs,
but they make up for this paucity by having a large number of individual sources. These galaxies were selected for observation because they are bright and have a spectacular morphology in all known wavebands.
Of the seven galaxies, four have been published already in the literature. The X-ray observations of the remaining three galaxies are unpublished except that in one case the literature reports only the identification of the brightest source with an interloper AGN.  We have analyzed their Chandra data, and list here the relevant findings for point sources (see Sect.~\ref{galaxies}).
To construct the XLF we will consider only sources along the ring.
Below we list and describe all the RiGs considered in this paper.

\subsection{Individual galaxies}
\label{galaxies}

Table~\ref{Tab:gallist} lists the relevant information for all the galaxies: name, position, Galactic line of sight absorption (N$_{\rm H}$) as used in this paper, distance in Mpc (as used in this paper), optical diameter (D25 from the NASA/IPAC Extragalactic Database - NED), and the scale at the distances of the galaxies. We describe below each galaxy in turn, concentrating mainly on the references and information about the X-ray analysis, the star formation rate (SFR) and the metallicity of the galaxies. 
We use as the primary SFR indicator the H${\alpha}$ derived value, which is most sensitive to the recent star formation. We expect the stellar population to be of young age (see below). This is also favouring the presence of massive OB stars as donors, which are the case for HMXB, although the issue is far from settled, as a few ULXs may have a late star as optical counterpart (e.g. IC 342 X1: \citet{Feng08}) and Main Sequence stars are favoured as donors in simulations, due to their longer life \citep[e.g.][]{Wiktorowicz17}. We list also, where available, other SFRs from the literature.

\begin{table*}
\begin{center}
\scriptsize
\caption{Observed ring galaxies.}
\label{Tab:gallist}
\begin{tabular}{lrrrrrrr}
\tableline\tableline
Name (Other name) & \multicolumn{2}{c}{Position} & N$_{\rm H}$  & Dist & Diam  & Scale\\
& \multicolumn{2}{c}{J2000} & 
[$10^{20}$ cm$^{-2}$]
& [Mpc]  & [$^{\prime}$] &  [kpc/$(^{\prime\prime})$]\\
\tableline
Cartwheel Galaxy (AM 0035$-$335)  & 00 37 41.1 &$-$33 42 59  &  19.0 &   122    & 1.5 &  0.566\\ 
NGC 922 (AM 0222$-$250)               & 02 25 04.4 &$-$24 47 17  &  1.62 &    48    & 2.6  & 0.239\\ 
Arp 147                               & 03 11 18.9 &$+$01 18 53  &  6.2  &   133    & 0.5 &   0.643\\ 
AM~0644$-$741 ({\scriptsize Lindsay-Shapley Ring})&06 43 06.1 &$-$74 13 35&10.0&  91.6   & 1.7 &   0.447\\ 
Arp 143                               & 07 46 53.6 &$+$39 01 10  &  5.01 &   57.1   & 3.0 &  0.283\\ 
Arp 148 (Mayall's object)             & 11 03 53.2 &$+$40 50 57  &  1.01 &  145.2   & 0.6 &  0.696\\ 
Arp 284 (NGC 7714/7715)               & 23 36 18.1 &$+$02 09 21  &  18.0 &    37    & 2.9 &  0.186\\ 
\tableline
\end{tabular}
\end{center}
\tablecomments{Data are from NED unless specified otherwise; N$_{\rm H}$ and distance from literature as in Table~\ref{Tab:journal}; scale is the plate scale at the galaxies' distance.}
\end{table*}
\subsubsection{\bf Cartwheel Galaxy (AM 0035$-$335)}

The \object[Cartwheel Galaxy]{Cartwheel} is the most famous object in the group. It has been studied extensively at all wavelengths. Several authors have studied this galaxy in great details both in the 
X-ray and other wavebands (e.g. \citealt{Wolter99,Gao03,WT04,Wolter06,Crivellari09,Pizzolato10} for the description of the X-ray data).
The Cartwheel contains 16 ULX \citep{WT04} and has a high SFR: \cite{WT04} find it consistent with the SFR $ \sim 20\,{}{\rm M}_\odot{}\,{\rm yr}^{-1}$  obtained from L$_{H\alpha}$ \citep{Appleton97}, while 
 \citet{Mayya05} derive a SFR $ = 18\,{}{\rm M}_\odot{}\,{\rm yr}^{-1}$ from radio data. The metallicity has been measured by \citet{FosburyHawarden77} at the position of the brightest HII regions in the ring, and the result is subsolar: Z = 0.14 Z$_{\odot}$, assuming Z$_{\odot}$ = 0.019 \citep{anders1989}.

The nuclear mass of the Cartwheel, derived from the rotation velocity in the ring, is M $= 4 \times 10^{11} {\rm M}_\odot{}\,$, while the ring contains stars and ionized gas for a total of $ {\rm M}_{star} \sim 2 \times 10^{10} {\rm M}_\odot{}\,$ and $ {\rm M}_{gas} \geq 1.2 \times 10^{8} {\rm M}_\odot{}\,$ \citep{FosburyHawarden77}.

CO is detected by ALMA only in the nucleus and inner ring \citep{Higdon15} for a total mass of 2.7 $\times 10^{9}$ M$_{\odot}$. The kinematic of the gas implies an age of the inner ring of $\sim 70 $ Myr. The outer ring appears dominated by an atomic phase of the gas, with a very little content of H$_{2}$. 
The age of the outer ring is estimated to be $\sim 440$ Myr (\citealt{Higdon95, Higdon96}).

\begin{table*}
\begin{center}
\caption{Galaxy properties from the literature}
\label{Tab:galprop}
\begin{tabular}{lrllll}
\tableline\tableline
Name    & M$_K$ & B & SFR & Z\tablenotemark{a}  & Ref\tablenotemark{b}  (SFR, Z)\\
 &  & & M$_{\odot}$/yr  & Z$_{\odot}$  & \\ 
\tableline
Cartwheel && 14.82\tablenotemark{c}  & 20 & 0.14 & 1, 2 \\ 
NGC 922        & -23.21 & 12.32 & 8.0 & 0.5 - 1. & 3, 3 \\ 
Arp 147       & -22.88 &15.99$\pm$0.05& 4.1 & 0.19 - 0.40 & 4, 5 \\
AM~0644$-$741 &&    11.0$\pm$0.4 &2.6 &  0.45 & 6, 7 \\
Arp 143       & -23.86 &13.30$\pm$0.08& 2.3 & 0.44 - 0.71 & 4, 8 \\ 
Arp 148       & -23.23 &16.04$\pm$0.03& 2.5 && 4, --  \\
Arp 284       &&& 4.0 & 0.19 - 0.38& 9,  10  \\ 
\tableline

\end{tabular}
\end{center}
\tablenotetext{a}{Assuming Z$_{\odot}$ = 0.02, 12+log(O/H)$_{\odot}$ = 8.92}
\tablenotetext{b}{Reference list: 
1: H$_{\alpha}$ data: \cite{Appleton97}; 
2: \cite{FosburyHawarden77}; 
3: H$_{\alpha}$ data \cite{Wong06}; 
4: H$_{\alpha}$ data: \cite{Romano08}; 
5: from log(NII/H$_{\alpha}$) in different ring positions \citep{Fogarty11};
6: H$_{\alpha}$ data \cite{HigdonWallin97};
7: \cite{Mapelli09};
8: \cite{Higdon97} citing \citet{Jeske86};
9: H$_{\alpha}$ data \cite{Schmitt06};
10: \cite{Garcia-Vargas97}.
}
\tablenotetext{c}{B-Magnitude of the ring}

\tablecomments{M$_K$ and B are from NED, SFR and Z from References in column 6.}

\end{table*}

In this paper we use the results of the X-ray analysis as presented in \cite{WT04}. 

\subsubsection{\bf NGC~922 (AM 0222$-$250)}

\object[NGC 922]{NGC~922} has been  described as a RiG only recently \citep{Wong06}. It contains 7 ULXs and has a SFR = 8$\,{}{\rm M}_\odot{}\,{\rm yr}^{-1}$ (\citealt{Prestwich12, Wong06}) derived from $H_{\alpha}$ measures, which we use hereafter. 

The ring has a distorted appearance that probably hindered the classification of the galaxy. However the high SFR and high number of ULXs are consistent with it being a collisional RiG. Metallicity has been published only for the nucleus of the galaxy \citep{Wong06} and is consistent with solar  (following \citealt{Prestwich12}, we adopt $Z\sim{}0.75$ Z$_\odot$). Optical spectra taken at optical knots positions suggest a smaller value for metallicity in the ring (Emanuele Ripamonti, private communication).
The gravitational interaction of the bullet galaxy (S2) with NGC 922 has likely happened about 330 Myr ago \citep{Wong06} with a slightly off-center impact. \citet{Pellerin10} find young stellar clusters in the ring and nuclear region, while the plume is made of older stars, consistent with the impact scheme. 

In this paper we use the results of the X-ray analysis as presented in \cite{Prestwich12}. 

\subsubsection{\bf Arp~147} 

\object[Arp147]{Arp~147} lives in one of the most diverse collections of celestial objects in a small patch of the sky: in the close vicinity of the RiGs there are the possible intruder galaxy, a bright star and an X-ray emitting AGN. All within less than 1$^{\prime}$.   \citet{Rappaport10} have analyzed this galaxy and list 9 ULXs and a SFR of $7\,{}{\rm M}_\odot{}\,{\rm yr}^{-1}$. 

A SFR of $4.1\,{}{\rm M}_\odot{}\,{\rm yr}^{-1}$ from H${\alpha}$ and of $8.6\,{}{\rm M}_\odot{}\,{\rm yr}^{-1}$ from FIR is derived by \cite{Romano08}. We use the H${\alpha}$ derived value. Metallicity has been measured via the N2 ratio defined as N2 = log(NII/H$_{\alpha}$) in different ring quadrants and ranges between Z = $0.19 - 0.40$ Z$_{\odot}$ \citep{Fogarty11}.

Less than $\sim{}50-80$ Myr have elapsed since the collision, as indicated by N-body simulations (\citealt{Gerber96, Mapelli12}).
In this paper we use the results of the X-ray analysis as presented in \cite{Rappaport10}. 

\subsubsection{\bf AM~0644$-$741 (Lindsay-Shapley Ring) }  

\object[AM~0644-741]{AM~0644$-$741} (see figure 1) shows an almost complete ring, and a partial second, more internal ring. It is probably the galaxy most similar to the Cartwheel in morphology, except for the presence of an active nuclear source. 

The brightest source in AM0644-741(\#3) has been observed spectroscopically by \citet{Heida13} and it has been identified with an AGN at z=1.40. The source is not positionally coincident with the ring and therefore does not belong to the sample which we use for the XLF.

In this paper we use a SFR of 2.6$\,{}{\rm M}_\odot{}\,{\rm yr}^{-1}$ derived from H$\alpha$ by \citet{HigdonWallin97} and sub-solar metallicity of  Z = 0.45 $_{\odot}$  \citep{Mapelli09}. However we are aware that  
\citet{Higdon11} 
derive a SFR = 11.2$\pm$0.4 M$_{\odot}$ yr$^{-1}$ from L$_{H\alpha}$ and the less favored SFR = 17.6$\pm$0.9 M$_{\odot}$ yr$^{-1}$ from radio measures, in addition to a metallicity in excess of solar with no significant azimuthal variations. 

The Chandra X-ray observation is not published so we downloaded it from the archive and  analyzed it as described below (see Setc.~\ref{analysis}). All 9 detected point sources are listed in Table~\ref{Tab:sourcelist}. Their luminosities are all above the ULX limit if associated to the galaxy. The list includes one source associated to the nucleus. In addition to \#3, we will exclude this source from the XLF computation. 

From the total detected net counts in the area of the galaxy (excluding the interloper AGN and the nuclear source) of 593.0$\pm$37.0 we measure a total unabsorbed luminosity associated with the galaxy of  L$_X (0.5-10 {\rm keV}) = 1.78 \times 10^{41}$ erg/s. We assumed a spectral shape that includes both a ``point-sources" spectrum with fixed $\Gamma =1.7$  and a thermal plasma ({\em mekal} in XSPEC) component to describe the diffuse gas, both absorbed by a fixed N$_{\rm H} = 0.1 \times 10^{22}$ cm$^{-2}$. The fraction accounted for by the thermal component corresponds to an unabsorbed L$_X (0.5-10 {\rm keV}) = 9.74 \times 10^{39} $ erg/s with kT = 0.08 ($<0.33$) keV.  
We extracted the counts for all detected point sources (178.1 counts in total) and derived a mean spectral shape of N$_{\rm H}$ = 0.22 [0.01-0.43]$\times 10^{22}$ cm$^{-2}$ and $\Gamma $ = 1.55 [1.18-1.99] for a total unabsorbed L$_X (0.5-10 $keV$) = 2.6 \times 10^{40} $erg/s. 
Most of the X-ray luminosity is in unresolved sources.

\begin{figure}[h]
\includegraphics[width=9.cm]{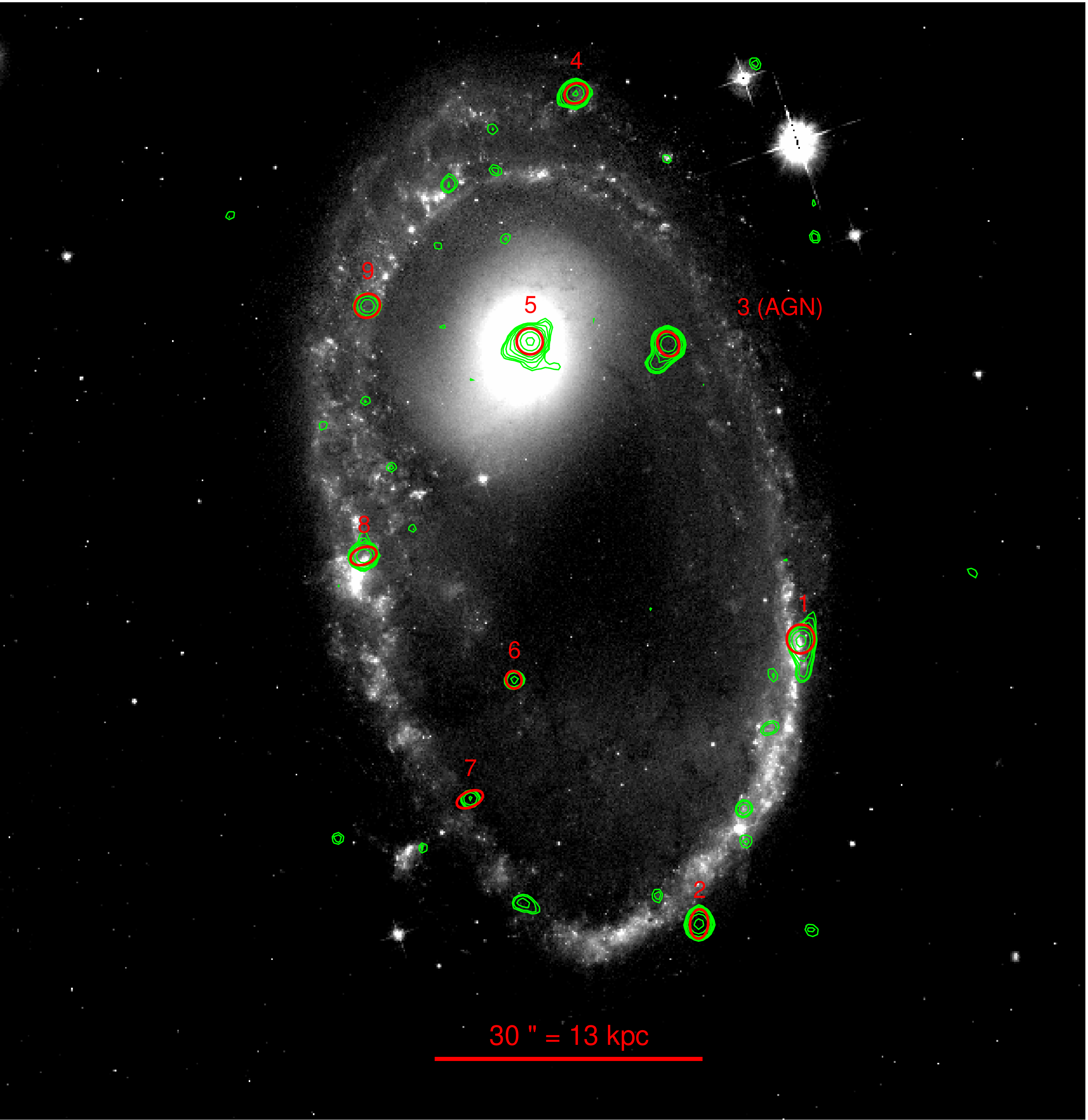}
\caption{\small AM0644$-$741. HST ACS WFC1 F555W image and superimposed the detected X-ray sources (in red) and X-ray contours (in green) from Chandra data. The scale of the figure is given in the red bar. \label{Am0644} }
\end{figure}

\begin{table*}
\begin{center}
\caption{Journal of Chandra observations}
\label{Tab:journal}
\begin{tabular}{lrlrr}
\tableline\tableline
Name  &OBSID & Date  & Time  & Ref. for X-rays  \\
   &  & & $\times 10^3$ sec  & \\
\tableline
Cartwheel Galaxy &2019  & 2001-may-26 & 76.15 & \cite{WT04}\\  
\hline
NGC 922          &10563 &2009-mar-05 & 29.74  & \cite{Prestwich12} \\ 
NGC 922          &10564 &2009-oct-02 & 10.02  & -- \\ 
\hline
Arp 147          &11280 &2009-sep-13 & 24.53  & \cite{Rappaport10} \\  
Arp 147          &11887 &2009-sep-15 & 18.04  & -- \\   
\hline
AM~0644$-$741    &3969  &2003-nov-17 & 39.40  & this paper\\  
\hline
Arp 143          &14906 &2012-dec-10 & 39.00  & this paper \\ 
\hline
Arp 148          &12977 &2011-feb-07 & 52.34  & this paper\\ 
\hline
Arp 284        & 4800 & 2004-jan-24 & 59.00   & \cite{Smith05}  \\ 
\tableline
\end{tabular}
\end{center}
\end{table*}
\subsubsection{\bf Arp~143}

\object[Arp 143]{Arp~143} (see figure 2) has a complex morphology. It consists of the galaxies NGC 2444 and NGC 2445. The Galex image clearly shows the young stellar population in the ring \citep{beirao09}. HI and CO are detected in the galaxy: HI in the ring and CO clustered around the nucleus \citep{Higdon97} together with a number of HII regions. A long ($\sim 100 $ kpc) HI tail (\citealt{Appleton87}) might represent the leftover of previous, perhaps multiple encounters. The kinematics of the ring implies an age of the ring of 60$\pm$15 Myr, about 20\% of the Cartwheel, and the N-body simulations of \citet{Gerber96}  imply that time elapsed since the collision is less than 80 Myr.
About 30 Myr have elapsed since the encounter, assuming NGC~2444 is the bullet galaxy and based on the dynamics and distance of the two galaxies (from \citealt{Jeske86} referred to in \citet{Higdon95}). The total molecular gas content is estimated to be H$_2$=4-24$ \times 10^9  M_{\odot}$  \citep{Higdon95}.

\citet{Romano08} find a SFR = $2.30\, {\rm M}_\odot{}\,{\rm yr}^{-1}$ from H$_{\alpha}$ and a SFR = $5.6\, {\rm M}_\odot{}\,{\rm yr}^{-1}$ from FIR. We use the H$_{\alpha}$ derived SFR and a sub-solar metallicity from \citet{Jeske86}, cited by \citet{Higdon97} as Z = $0.44 - 0.71$ Z$_{\odot}$.

The Chandra X-ray observation is not published so we downloaded it from the archive and  analyzed it as described below (see Sect.~\ref{analysis}). All 14 detected point sources are listed in Table~\ref{Tab:sourcelist}.
Both nuclei are detected in the galaxies forming Arp~143, plus an interloper Sy~1 at z=0.6155313 \citep{Abolfathi17}.
Two sources have a radio counterpart including the nucleus of NGC 2445. 
A couple more of X-ray sources (noted in Table~\ref{Tab:sourcelist}) are not consistent with the ring position and will not be used to derive the XLF. 

\begin{figure}[h]
\includegraphics[width=9.cm]{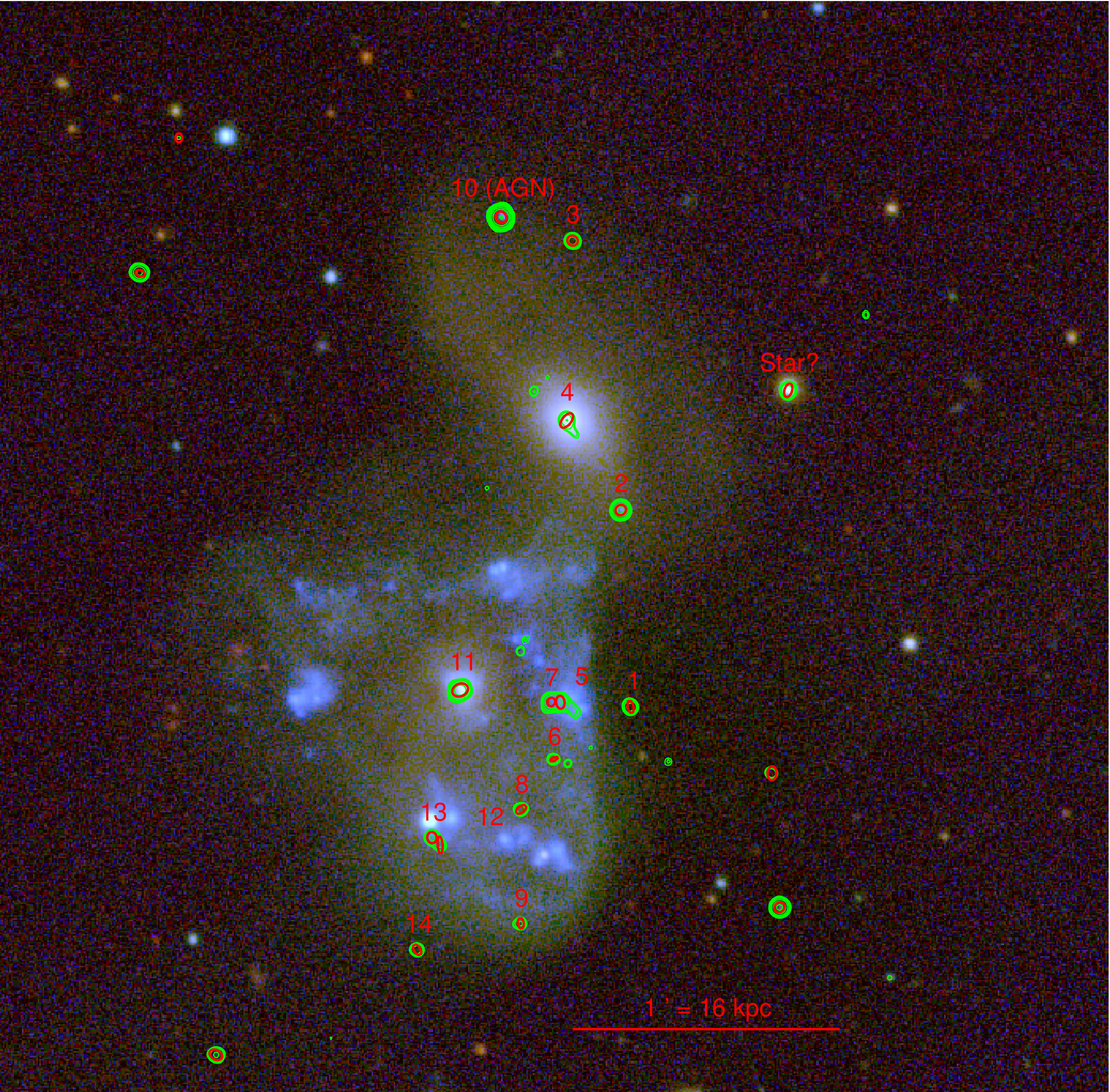}
\caption{\small Arp 143. SDSS-III image (RGB representation with SDSS bands z, g, u respectively) and superimposed the detected X-ray sources (in red) and X-ray contours (in green) from Chandra data. The six sources outside the boundaries of the galaxy pair were not considered in this paper. The scale of the figure is given in the red bar. \label{Arp143} }
\end{figure}

From the total detected net counts in the area of the galaxy (excluding the Sy1 and the two nuclear sources) of 490.6$\pm$39.6 we measure a total unabsorbed luminosity associated with the galaxy of L$_X (0.5-10$ keV$) = 4.81 \times 10^{40} $erg/s. We assumed a spectral shape with fixed N$_{\rm H}$= $0.05 \times 10^{22}$ cm$^{-2}$ that includes a "point-sources" spectrum with fixed $\Gamma =1.7$ and a thermal plasma ({\em mekal} in XSPEC) component to describe the diffuse gas.  The fraction accounted for by the thermal component corresponds to an unabsorbed L$_X (0.5-10$ keV$) = 7.99 \times 10^{39} $erg/s with kT = 0.50 (0.32-0.66) keV.  
We extracted the counts for all detected point sources (247.1 counts in total) and derived a mean spectral shape of N$_{\rm H}$= 0.074 [$<0.187$]$\times 10^{22}$ cm$^{-2}$ and $\Gamma $ = 1.57 [1.26-1.94] for a total unabsorbed L$_X (0.5-10$ keV$) = 2.6 \times 10^{40} $erg/s.

\subsubsection{\bf Arp~148 (Mayall's object)} 

\object[Arp 148]{Arp~148} (see figure 3) is a beautiful example of an interacting pair. We detect only two sources in the ring, beside the foreground member of the pair, which is also a radio source. However the limiting flux of the Chandra observation corresponds to much higher luminosity of L$_X  = 5 \times 10^{39} $erg/s than the other galaxies, given the much larger distance and comparatively short exposure time. 

The galaxy lies in fact at D=145 Mpc \citep{Strauss92} with a low galactic line of sight absorption of N$_{\rm H}= 1.04\times 10^{20}$ cm$^{-2}$. 
The source was observed by Chandra on 7 Feb 2011 for a total of 52.34 ksec. The X-ray observation is not published so we downloaded it from the archive and  analyzed it as described below (see Sect.~\ref{analysis}). All the detected point sources are listed in Table~\ref{Tab:sourcelist}.
In our analysis we used a SFR = $2.5\,{}{\rm M}_\odot{}\,{\rm yr}^{-1}$ from H$_{\alpha}$ (\citealt{Romano08}). A SFR = $65.06\,{}{\rm M}_\odot{}\,{\rm yr}^{-1}$ is derived from the far infrared (\citealt{Romano08}), however this measurement could be affected, given the low resolution of the IRAS data, by confusion with the companion galaxy, which could host an AGN, being a radio emitter. 
The metallicity is not known.
\citet{Romano08} estimate an elapsed time since collision of less than 80 Myr, based on N-body simulations of \cite{Gerber96}.

The total unabsorbed luminosity associated with the ring galaxy, computed with a spectral shape with fixed N$_H = 0.01 \times 10^{22}$ cm$^{-2}$ 
that includes a point sources spectrum with fixed $\Gamma =1.7$ and a thermal plasma ({\em mekal} in XSPEC) component to describe the diffuse gas, is of  L$_X (0.5-10 keV) = 3.87 \times 10^{41} $erg/s given the total detected net counts in the area of the ring galaxy (excluding the interloper)  of 73.6$\pm$13.3.  The fraction accounted for by the thermal component corresponds to an unabsorbed L$_X (0.5-10 keV) = 1.28 \times 10^{41} $erg/s with kT = 0.23 (0.17-0.31) keV.  However the small number of counts does not allow a very reliable fit. 
As a last step, we extracted the counts for all detected point sources (34.0 counts in total) and derived a mean spectral shape of N$_{\rm H}^{fixed} = 0.01 \times 10^{22}$ cm$^{-2}$ and $\Gamma $ = 1.63 [1.14-2.14] for a total unabsorbed L$_X$ is L$_X (0.5-10 keV) = 1.5 \times 10^{40} $erg/s .   

As regards the intruder galaxy, it is detected with total net counts = 384.0$\pm$ 20.9. We fit the extracted spectrum with a power law to represent both the nucleus and possible unresolved point sources, plus a thermal plasma ({\em mekal} in XSPEC) component to account for the diffuse plasma. Both components are necessary to the fit. We fix the absorbing hydrogen column density to the Galactic value of N$_{\rm H} = 0.01 \times 10^{22} {\rm cm^{-2}}$. The resulting $\Gamma$ = 1.63 [1.25-1.96] is perfectly consistent with both unresolved point sources or the presence of a nucleus, while the kT = 0.66 [0.58-0.74]keV is also similar to temperatures found for the gas in the other RiGs and for local galaxies in general \citep{bogdan11}. The unabsorbed flux is f$_X (0.5-10 {\rm keV}) =  5.02 \times 10^{-14}$ erg/cm$^{2}$/s for a luminosity L$_X (0.5-10 {\rm keV}) = 1.39 \times 10^{41} $erg/s where the two components contribute as follows: the power law L$_X (0.5-10 keV) = 9.84 \times 10^{40} $erg/s and the diffuse gas L$_X (0.5-10 {\ keV}) = 4.08 \times 10^{40} $erg/s.

\begin{figure}[h]
\includegraphics[width=9.cm]{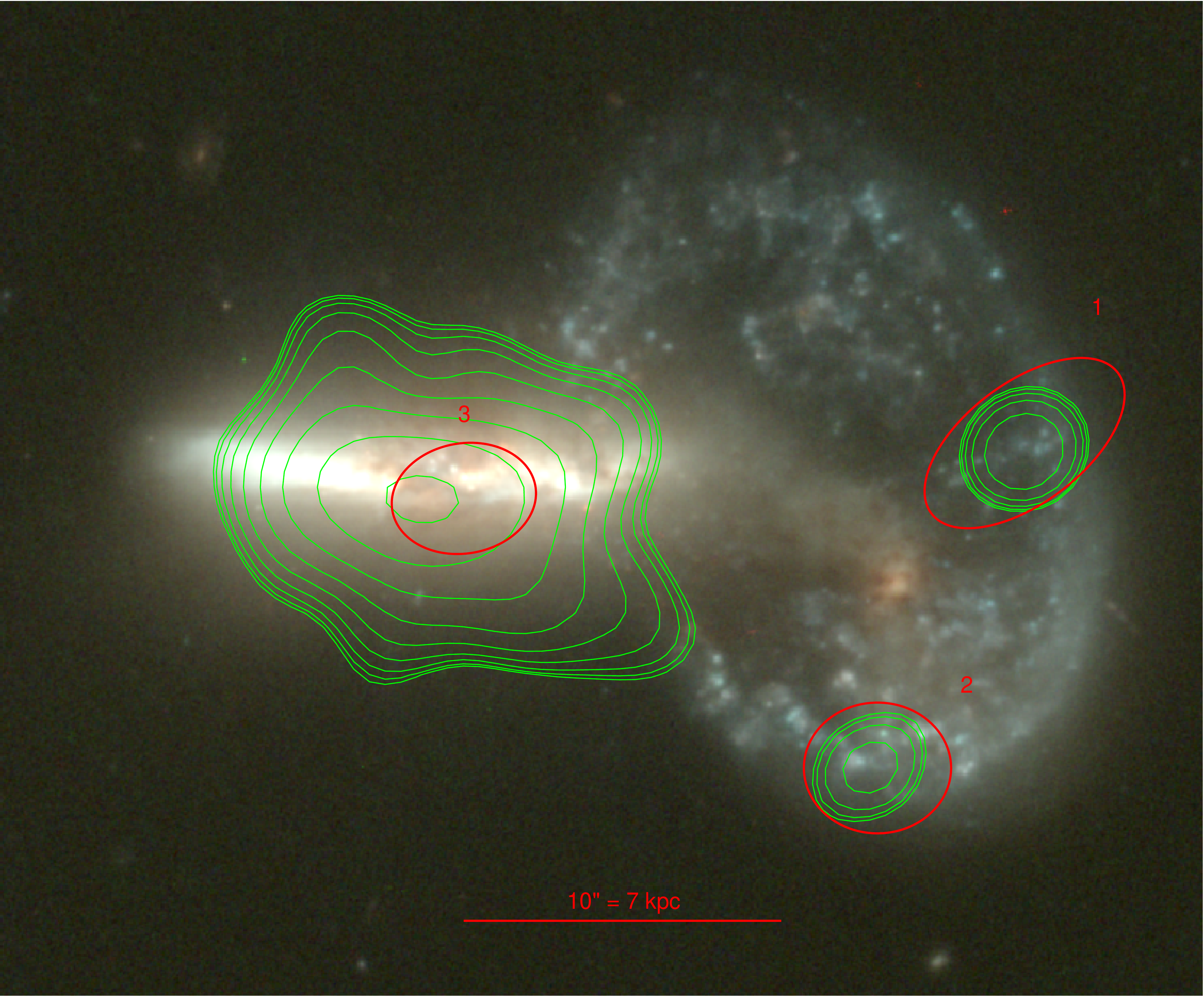}
\caption{\small Arp 148. Hubble Heritage image (WFP2 F814W/F555W/F450W) and superimposed the detected X-ray sources (in red) and X-ray contours (in green) from Chandra data. The scale of the figure is given in the red bar. \label{Arp148} }
\end{figure}

\subsubsection{\bf Arp~284 (NGC 7714/7715)} 

\object[Arp 284]{Arp~284} is the interacting pair NGC~7714 and NGC~7715, with many structures, for instance bridges between the two galaxies and tails. The galaxy contains 9 ULXs and has a SFR = 7$\,{}{\rm M}_\odot{}\,{\rm yr}^{-1}$ (\citealt{Smith05}). 
While \citet{Lancon01} finds a SFR $\sim 1\,{}{\rm M}_\odot{}\,{\rm yr}^{-1}$ from complex SED modeling, \citet{Schmitt06} find a value of SFR = 4.0$\,{}{\rm M}_\odot{}\,{\rm yr}^{-1}$ from H${\alpha}$ data and a SED correction. We use this last H${\alpha}$ derived value. The metallicity has been measured by \citet{Garcia-Vargas97} to be Z = $0.19-0.38$ Z$_{\odot}$. 
We use only the sources in the ring, as listed in Table 5,
to construct the XLF.

In this paper we use the results of the X-ray analysis as presented in \cite{Smith05}. 

\begin{table*}
\begin{center}
\scriptsize
\caption{X-ray detected sources in AM0644-741, Arp~148 and Arp~143.}
\label{Tab:sourcelist}
\begin{tabular}{lrrrrrrrl}

\tableline\tableline
Source \# &  \multicolumn{2}{c}{Pos} & Counts & f$_{\rm X}^{[0.5-10 {\rm keV}]}$ & L$_{\rm X}^{[0.5-10 {\rm keV}]}$ & L$_{\rm X}^{[2-10 {\rm keV}]}$ & XLF Corr & NOTES\\
  & \multicolumn{2}{c}{(FK5)}  && (a) & (b) & (b) & (c)\\
\tableline
AM0644x1 & 06 42 58.62 & -74 14 44.1 &  28.9$\pm$ 5.5  & 7.95$\pm$2.1 & 8.0 & 5.1 & 1.0 \\ 
AM0644x2 & 06 43 01.40 & -74 15 16.1 &  46.3$\pm$ 6.9  &12.6$\pm$2.8  & 12.6& 8.1 & 1.0 \\
AM0644x3 & 06 43 02.27 & -74 14 11.1 & 107.3$\pm$ 10.4 &28.5$\pm$5.1  & 28.6& 18.3 & -- & AGN in \cite{Heida13}\\
AM0644x4 & 06 43 04.81 & -74 13 43.1 &  36.4$\pm$ 6.1  & 9.8$\pm$2.7  & 9.8 & 6.3  & 1.0 \\ 
AM0644x5 & 06 43 06.08 & -74 14 10.9 &  85.1$\pm$ 9.3  &22.3$\pm$4.1  & 22.4& 14.3 & -- & Nucleus\\
AM0644x6 & 06 43 06.51 & -74 14 48.7 &   7.6$\pm$ 2.8  & 2.1$\pm$1.3  & 2.1 & 1.3  & 1.67 &  \\ 
AM0644x7 & 06 43 07.71 & -74 15 02.1 &   8.4$\pm$ 3.0  & 2.3$\pm$1.5  & 2.3 & 1.4  & 1.33 &  \\
AM0644x8 & 06 43 10.54 & -74 14 06.9 &  19.2$\pm$ 4.5  & 5.1$\pm$1.7  & 5.1 &  3.3 & 1.05 \\ 
AM0644x9 & 06 43 10.64 & -74 14 34.9 &  31.4$\pm$ 5.7  & 8.5$\pm$2.2 & 8.5 &   5.4 & 1.0 \\ 
\tableline
Gal & 06 43 25.50 & -74 15 26.2 & 30.41$\pm$6.78 & 6.96$\pm$2.6 & 7.0 & 4.44 & -- & possible bullet? \cite{Few82}\\
\tableline
\tableline
Arp148x1 & 11 3 52.35 & +40 51 1.4 & 18.4$\pm$ 4.4 &  3.5$\pm$0.9 &  8.8 & 5.7  & 1.0 &  \\
Arp148x2 & 11 3 52.74 & +40 50 51.3  & 14.8$\pm$ 4.0 &   2.8$\pm$0.9   & 7.1 & 4.6 & 1.0 & \\
Arp148x3 & 11 3 53.99 & +40 51 0.1  &139.2$\pm$ 11.9  &  25.8$\pm$4.0  &  65.1 & 42.1 & --& Nucleus/Galaxy \\
\tableline
\tableline
Star? & 7 46 48.81 & +39 02 01.7   &  28.7$\pm$5.39  &   8.6$\pm$2.8 & -- & -- & -- & not related\\
\tableline 
Arp143x1  & 7 46 51.85 & +39 00 50.6  & 21.8$\pm$4.7  &  6.9$\pm$2.6 & 2.7 & 1.7& 1.0 & \\
Arp143x2  & 7 46 52.04 & +39 01 34.8  & 83.7$\pm$9.2  & 25.0$\pm$4.7 & 9.8 & 63  & --& bridge? \\
Arp143x3  & 7 46 52.97 & +39 02 35.4  & 23.7$\pm$4.9  &  7.3$\pm$2.3  & 2.9 & 1.9 &-- &not on ring (N2444) \\
Arp143x4  & 7 46 53.08 & +39 01 54.9  &  16.5$\pm$4.1 & 4.8$\pm$1.9 &  1.9 & 1.2 & --& Nucleus N2444 \\
Arp143x5  & 7 46 53.20 & +39 00 51.6  & 32.7$\pm$5.7  &  9.6$\pm$2.5 & 3.8 & 2.5 & 1.0 &radio source\\
Arp143x6  & 7 46 53.32 & +39 00 38.8  &  7.9$\pm$2.8  &  4.0$\pm$2.4 & 1.6 & 1.0 & 1.25 & \\
Arp143x7  & 7 46 53.38 & +39 00 51.7  & 35.8$\pm$6.0  & 10.9$\pm$2.7 & 4.3 & 2.8 & 1.0 & \\
Arp143x8  & 7 46 53.96 & +39 00 27.5  & 11.8$\pm$3.5  &  3.6$\pm$1.8 & 1.4 & 0.9 & 1.10& \\
Arp143x9  & 7 46 53.97 & +39 00 01.9  &  9.8$\pm$3.2  &  3.1$\pm$1.5 & 1.2 & 0.8 & 1.18& \\
Arp143x10 & 7 46 54.35 & +39 02 40.6  & 669.6$\pm$25.9&226.0$\pm$14.0 & $2.8 \times 10^5$ & 1.8$\times 10^5$  & -- & Sy1 z=0.6155313 (SDSS) \\
Arp143x11 & 7 46 55.14 & +39 00 54.4  & 66.4$\pm$8.2 &  19.2$\pm$3.9 & 7.5 & 4.8 & -- & Nucl.N2445+radio source\\
Arp143x12 & 7 46 55.52 & +39 00 19.5  & 10.7$\pm$3.3 &   3.7$\pm$1.7 & 1.4 & 0.9 & 1.10& \\
Arp143x13 & 7 46 55.68 & +39 00 21.2  & 18.7$\pm$4.4  &  5.6$\pm$2.1 & 2.2 & 1.4 & 1.02 & \\
Arp143x14 & 7 46 55.96 & +38 59 56.0  & 12.8$\pm$3.6  &  4.2$\pm$2.0 & 1.6 & 1.0 & 1.09 & \\
\tableline

\end{tabular}
\end{center}
\tablenotetext{a}{Unabsorbed flux in units of  $10^{-15}$ erg/cm$^{2}$/sec }
\tablenotetext{b}{Unabsorbed luminosity in units of $10^{39}$ erg/sec }
\tablenotetext{c}{Multiplicative correction as derived from \citet{KimFabbiano03} }
\tablecomments{Fluxes (0.5-10.0 keV) are computed by \textsc{ciao/srcflux} assuming N$_{\rm H}$ as in Table 1 and $\Gamma$ = 1.7. Quoted errors are all 90\%.}
\end{table*}

\subsection{Data Analysis}
\label{analysis}

Chandra observations for all ring galaxies were 
performed with the Advanced CCD Imaging Spectrometer S Array (ACIS-S; 
\citealt{garmire03}), and their log is reported in Table~\ref{Tab:journal}. 
For each source we list the Observation Id (ObsID), the date of the observation together with the total live time and the references for the published data for four of the galaxies. We refer to the original papers for a detailed description of the data analysis.
Three of the galaxies are presented here for the first time (see Sect.\ref{galaxies} and Table~\ref{Tab:sourcelist}).
The ACIS detector was operated in the standard timed exposure full-frame mode, 
with FAINT (Arp 143) or VERY FAINT telemetry format. In all cases the targets were positioned on the back-illuminated chip S3 (CCD7). 
The data were reprocessed with the Chandra Interactive Analysis of Observations 
software package (\textsc{ciao}, version 4.7 and 4.8, \citealt{fruscione06}) 
and \textsc{caldb} version 4.6/4.7\footnote{http://cxc.harvard.edu/ciao/} (we note here that no significant changes have occurred in \textsc{ciao} or the \textsc{caldb} that could affect our analysis).   
We followed standard data reduction and analysis procedures: after downloading the data from the public archive, we run the \textsc{chandra\_repro} tool to perform all the recommended data processing steps, then the \textsc{fluximage} tool to generate exposure-corrected images. 
Source detection was performed with the \textsc{wavedetect} tool on an image of the S3 chip generated by  \textsc{fluximage} in the ''broad'' energy range (0.5-7 keV) after excluding low-exposure regions at the edge of the chips (the \textsc{expmapthresh} parameter in  \textsc{fluximage}).
The detected sources are visible in figures 1, 2 and 3 as positional error ellipses. 
We computed count rates and fluxes for all the detected sources using the \textsc{srcflux} tool and assuming a power law model with photon index $\Gamma=1.7$ absorbed by an Hydrogen column density of $N_H$ as listed in table 1 for each galaxy. 
This choice is appropriate for ULXs (e.g. \citet{Swartz11}) which are unresolved at the Chandra resolution. 
The instrumental responses (ARF and RMF) were calculated for each source, while the PSF contribution to both the source and background regions was calculated with the \textsc{arfcorr} method (where a simulated circularly symmetric PSF is used to estimate the aperture correction). 
Fluxes were calculated in the 0.5-10 keV and 2-10 keV energy bands for all the sources and are listed in Table~\ref{Tab:sourcelist}.  Luminosities were calculated assuming the galaxy distance from table~\ref{Tab:gallist}. The "XLFcorr" column contains the correction factors for each source as described in Sect.~\ref{xlf}

In Fig.~\ref{Am0644}, ~\ref{Arp143} and ~\ref{Arp148} we display the optical image of the galaxies AM0644$-$741, Arp 143 and Arp 148 and overplot the detected sources and X-ray contours from Chandra data.
The description of the main results for each galaxy is in Sect.~\ref{galaxies}

Table 5 lists all the sources from the literature. For sake of uniformity and with the aim of constructing the X-ray Luminosity Function we recalculated luminosities in the 0.5-10 keV band for all the ring sources from the literature. In all cases we assumed the published $N_H$ and a power law model with photon index $\Gamma=1.7$.

\section{X-ray Luminosity Function}
\label{xlf}

In order to study the ULXs as a class and to be able to compare the results with different models and with results from different samples, we construct the X-ray luminosity function (XLF) by incorporating the sources detected in all the rings. The purpose is to see if an encounter as violent as the splashdown of a bullet galaxy into another galaxy is so energetic as to create a different ''flavor'' of ULXs, or a different distribution in luminosity. RiGs might host the most extreme class of ULXs. We assume that the ring population represents a single burst of star formation, produced by the impact, and that the populations from the different galaxies are homogeneous and can be combined.
Luminosities are taken from Table 4 and Table 5, recalculated in the (0.5-10) keV band for uniformity with previous studies.
We have collected a total of 63 sources, of which 50 above the ULX limit (10$^{39}$ erg/s) .
There are 23 sources with  $L_X> 5 \times 10^{39}$ erg s$^{-1}$ and 8 with  $L_X> 1 \times 10^{40}$ erg s$^{-1}$.

In order to take into account the detection confusion due the presence of diffuse and unresolved emission linked to the galaxies, we have applied a correction to the efficiency of detection following \citet{KimFabbiano03}. We base our assessment on fig. 12 from \citet{KimFabbiano03} in which the detection probabilities as a function of background counts and source counts are plotted. The background includes both diffuse emission from
the galaxy and field background and is estimated for each galaxy 
by averaging the total counts in the ring region, after excluding detected
sources. We use the second panel, relative to a 2 arcmin off-axis,
since the galaxies are always entirely within this limit. We apply the interpolated
correction - listed in Tab.~\ref{Tab:sourcelist} and Table 5 - to all sources with less than 22 counts. This represents a conservative estimate.

We assess the probability of interlopers by measuring the number of sources
expected by chance in the area of the rings in which we run the detection algorithm.
We conservatively use an elliptical annulus that includes all of the optical extent of the ring. The density
of sources at the limiting fluxes for each galaxy is derived from \cite{Moretti03}. 
Since the LogN-logS is given in the (0.5-2 keV) band, we used the 0.5-2 keV flux limit for each field (computed along the galaxies' ring) derived with the same spectral hypothesis. The total expected number of interlopers is $<$ 1 for each galaxy at the relative detection limit and in total it is $\sim 1.1$.
Given these numbers, we disregard the interlopers and remove no further sources from the XLF. 

\begin{figure}[h]
\includegraphics[width=9.cm]{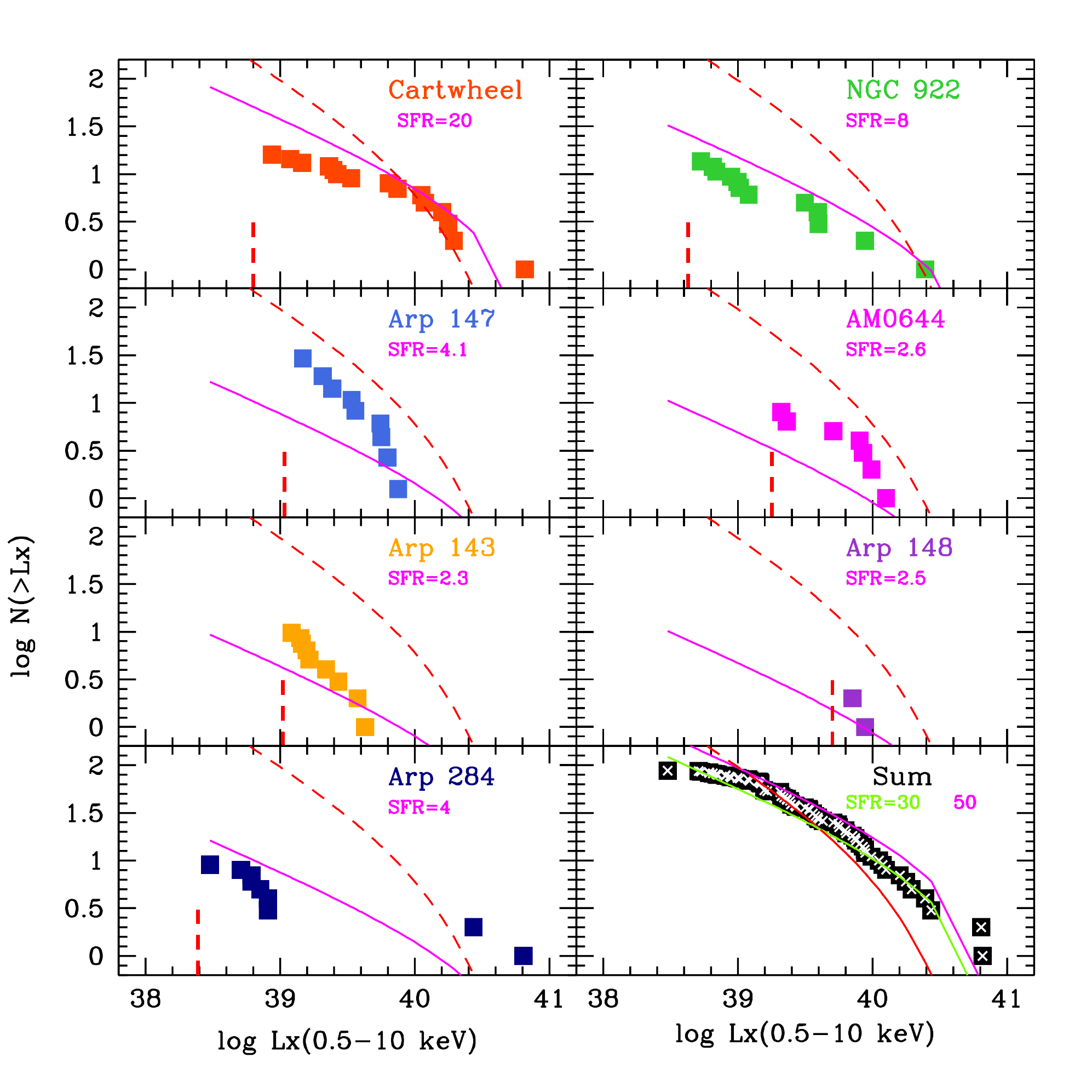}
\caption{\small The X-ray luminosity function (XLF) for the seven individual galaxies with upper limits (vertical dashed line) as in Tab.~\ref{Tab:ulim}.
The dashed red line is the model from \citet{Swartz11}.
The pink solid lines are the models of \cite{Grimm03}, normalized to the SFR from Table~\ref{Tab:galprop} as indicated in the panels (see text also). We stress that the uncertainty on the SFR measure is large and therefore the value for the single galaxy might not be representative.
\label{Fig:xlfSingle} }
\end{figure}

\begin{figure}[h]
\includegraphics[width=9.cm]{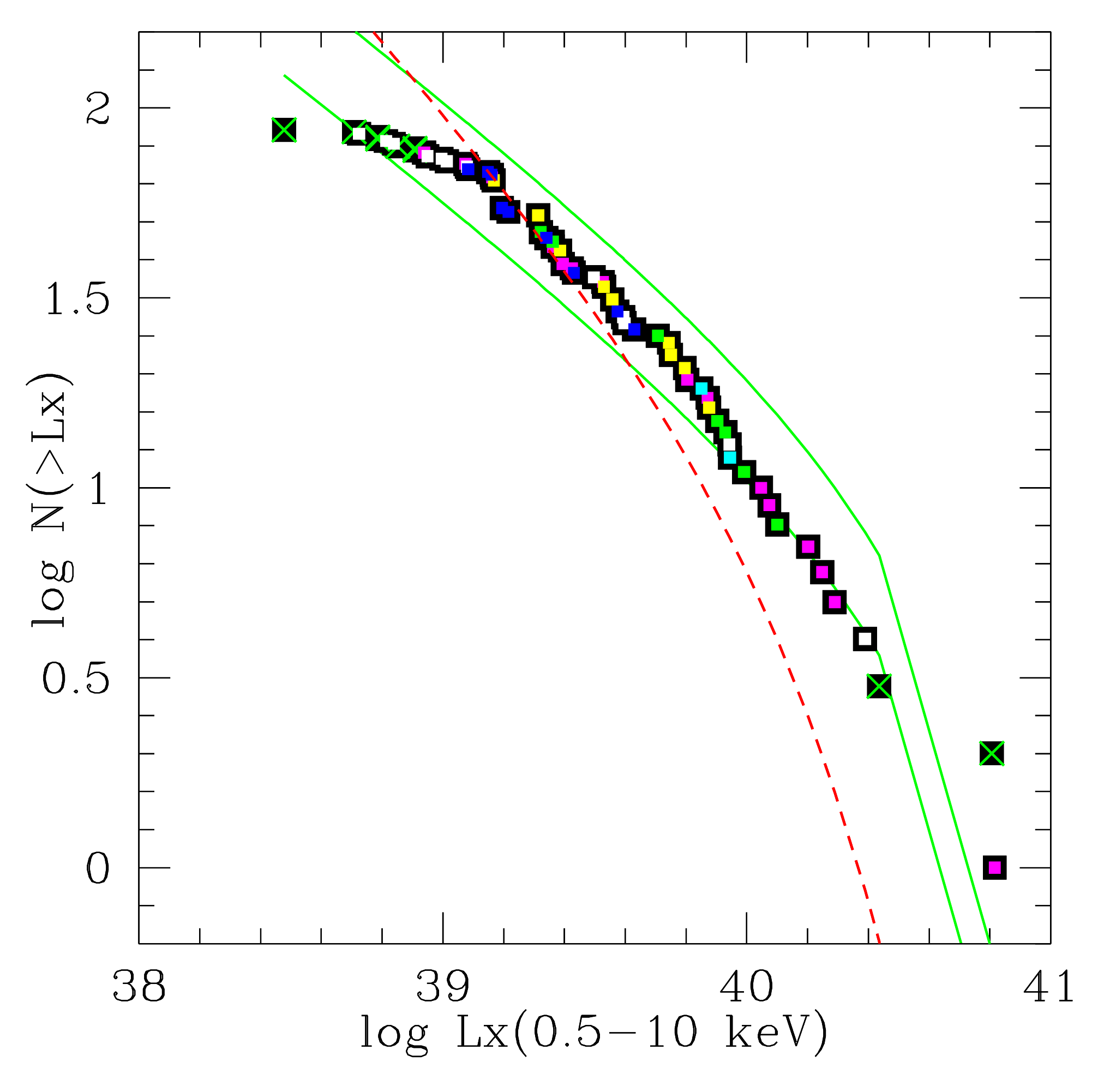}
\caption{\small The total X-ray luminosity function (XLF) derived from the 7  observed ring galaxies (see text for details). The green line is the model of \cite{Grimm03}, normalized to SFR $= 35$ ${\rm M}_\odot{}$ yr$^{-1}$ (lower curve), and to 50 ${\rm M}_\odot{}$ yr$^{-1}$ (upper curve). The sum of the SFR for the 7 galaxies from Table~\ref{Tab:galprop} is 43.5 ${\rm M}_\odot{}$ yr$^{-1}$. Sources with the same color are from the same galaxy. 
The dashed red line is the model from \citet{Swartz11} that corresponds to an effective SFR =  50 ${\rm M}_\odot{}$ yr$^{-1}$ .
\label{Fig:xlf} }
\end{figure}

We plot the resulting XLF in Fig.~\ref{Fig:xlfSingle} for each galaxy and in Fig.~\ref{Fig:xlf} for the sum of all sources, distinguishing sources from the same host galaxy by the same color and symbol.
We notice that, since not all observations have
the same limiting luminosity, the sampling at low L$_X$ is still not
complete, even after the correction for the higher background due to the galaxy contribution. 
This explains the flattening below L$_X \sim 10^{39}$ erg/s: since we are
interested mainly in the high luminosity end of the XLF, we do not consider this fact as a major problem at this time.

We have considered also two potential observational biases that could affect the final XLF. First, high absorbing columns of gas within the galaxies could obscure ULXs and lead to an underestimate of their numbers. For instance, \citet{Luangtip15} find that the high absorbing columns of 10$^{22}-10^{24}$ cm$^{-2}$ and dust content in luminous infrared galaxies might be a cause for the smaller number of ULXs per SFR unit they found. However, in RiGs the absorbing columns due to our Galaxy are in the 10$^{21}$ cm$^{-2}$ range at most, and fitted values in the ULX spectra (when statistics allows to perform a fit) are about at the same 10$^{21}$ cm$^{-2}$ level. Furthermore, the effect of a high absorbing column would be stronger for lower luminosity sources. We have sufficiently low detection limits, except for the case or Arp148, that this should not be a problem. The most significant cause for losing low flux sources would be the enhanced background due to the diffuse galactic emission, which we corrected for.
Second, the finite angular resolution may cause multiple sources to be detected as a single one, with an artificially higher luminosity, with the effect of flattening the high luminosity end of the XLF. If we take the Cartwheel as a reference, we know that most of ULXs are variable \citep{Crivellari09,Wolter06}, and therefore more likely to be single sources. A higher X-ray spatial resolution could indeed resolve a few ULXs. With the statistics in hand, however, we do not find any difference in the high L$_X$ end of the XLF by dividing the sample in near and far galaxies (most likely to be unresolved). The statistics however is not large, and therefore more numerous samples are needed to estimate if the effect is present at low levels.
  
We compare our results with previous work on XLFs.
In late-type galaxies, the occurrence and the luminosity of bright HMXBs (see e.g.
\citealt{Grimm03,Ranalli03,Mineo11})
and of  ULXs 
(e.g. \citealt{Mapelli09,Mapelli10}) 
are found to correlate with the SFR.

It was shown already with the XLF of the Cartwheel Galaxy \citep{WT04} and of NGC 2276 \citep{Wolter15} that the \cite{Grimm03} relation
linking the HMXB luminosity function with the SFR of the galaxy
gives consistent results.
\citet{Grimm03} use 14 galaxies with a total of 99 sources above  
$L_X> 2 \times 10^{38}$ erg s$^{-1}$ in the (2-10 keV) band. 
The \citet{Grimm03} function is normalized using the 
sum of the SFR values for the seven galaxies (SFR$_{Tot} = 43.5 \,{}{\rm M}_\odot{}/{\rm yr}$, see Sect.~\ref{galaxies}).  The luminosity is rescaled to our 0.5-10 keV band by using a power law slope of $\Gamma = 1.7$. 
Given the uncertainties in SFR measures, we plot for reference the \citet{Grimm03} function normalized to SFR = 35 M$_{\odot}$ yr$^{-1}$ and SFR = 50 M$_{\odot}$ yr$^{-1}$ which represent a lower and an upper envelope for the XLF, respectively.

\citet{Mineo12} selected a sample of 29 local galaxies (D $< 40$ Mpc) with SFR in the range 
0.1 to 15  ${\rm M}_\odot{}/{\rm yr}$ to discuss the connection with recent star formation activity and binary evolution. They point out that a few incorrect assumptions have been used to derive the \citet{Grimm03} functional form, however their relation between the SFR and the total $L_X$ is very similar to \citet{Grimm03}. It is of interest here to verify the relation with the total luminosity of HMXBs, given in their eq. (22): 
$L_{X} (\textrm{erg s}^{-1}) = 2.61 \times 10^{39} ({\rm SFR}/M_{\odot} yr^{-1})$. 
The total luminosity of the sum of the detected point sources is definitely consistent with this estimate, although a similar scatter in the $L_X$ vs. SFR relation is evident. This could be due either to large uncertainties in the determination of the correct SFR, or to some intrinsic physical effects.

A large sample of ULXs has been assembled by \cite{Swartz11} by searching about a hundred galaxies, up to D = 14.5 Mpc. The total
effective SFR is $ \sim 50 \,{}{\rm M}_\odot{}\,{}{\rm yr}^{-1}$. We plot their XLF curve in Figure~\ref{Fig:xlf} for reference as a dashed line, after rescaling the luminosity to the 0.5-10 keV band by using a power law slope of $\Gamma = 1.7$. The number of ULXs in the \cite{Swartz11} sample is 86 (to be compared with our 50) at L$_X \geq 10^{39}$ erg/s; 16 (to be compared with our 23) at L$_X \geq 5 \times 10^{39}$ erg/s; and 6 (to be compared with our 8) at L$_X \geq 10^{40}$ erg/s. At face value the RiGs XLF is flatter than the Swartz et al. one. However, 
a straightforward comparison would be incorrect, since we know that we have lost some sources at low fluxes.  Figure~\ref{Fig:xlf} shows a more significant comparison between the distribution corrected for the presence of diffuse emission and the \cite{Swartz11} fit.
In comparison with the \citet{Grimm03} curve we see an excess of only two sources at high luminosities, indicating that a higher cut-off L$_X$ might be more appropriate for the RiG XLF. The trend of RiGs XLF to be flatter with respect to \cite{Swartz11} is still present, indicating a larger incidence of high $L_X$ sources, consistent with the fact that the \citet{Swartz11} sample includes galaxies of all types including ellipticals, and therefore has a substantial population of low-mass X-ray binaries with a steeper XLF.
We think that the outliers effect merits further investigation even though the small number statistics definitely affects the results.

\section{Discussion}

We have presented the first XLF derived solely from ring galaxies and in particular from sources associated to the ring itself. We deem this a ``clean'' sample of ULXs produced from a single burst of recent star formation in a peculiar and energetic environment. 

There are additional X-ray sources in the galaxies' area, but their chance of being interlopers is larger than in the ring 
(see e.g.  \citet{Heida13, WT04,Wolter15})  otherwise 
and they could be due to a different formation mechanism or a different episode of star formation.

\begin{figure}[h]
\includegraphics[width=9.cm]{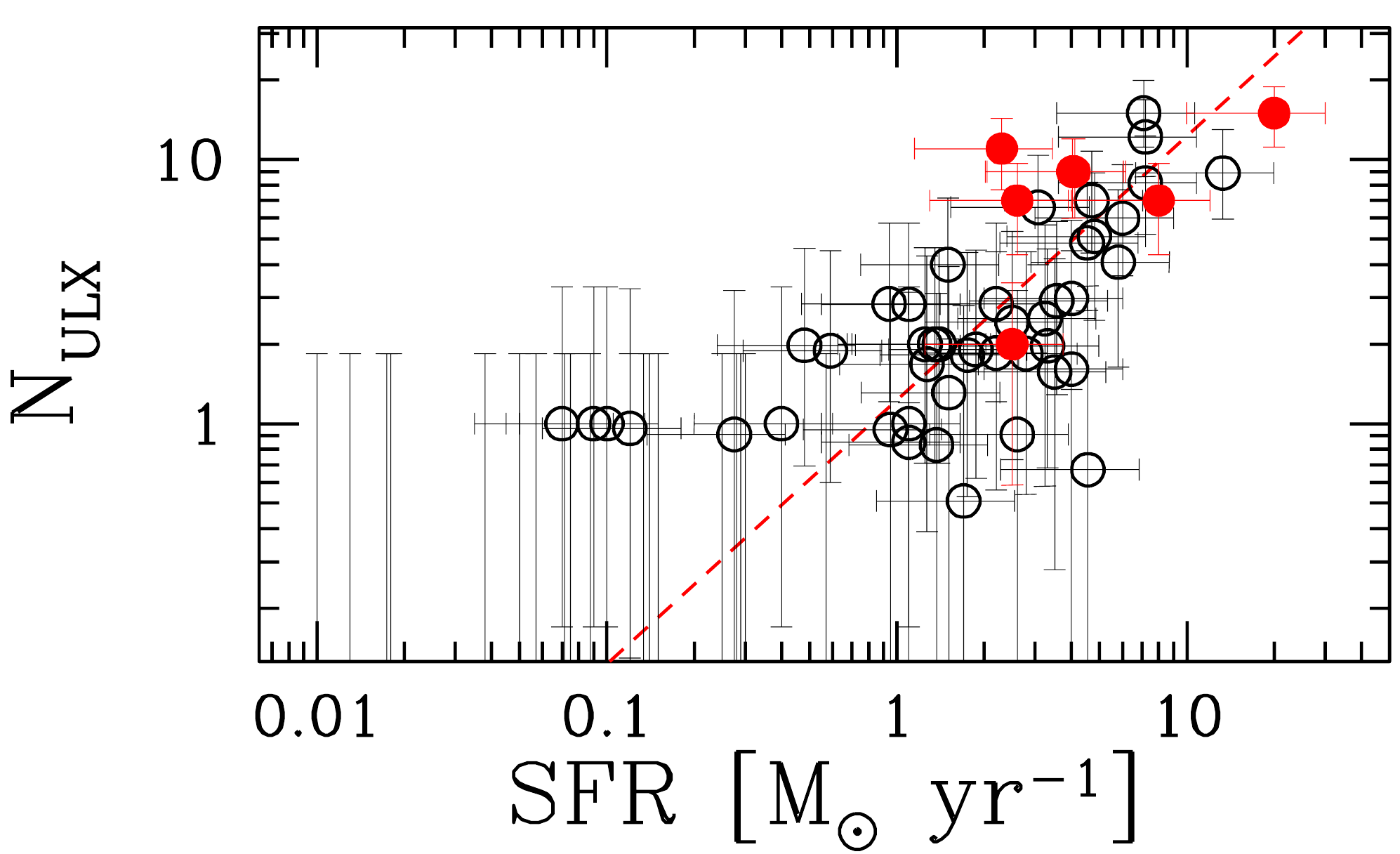}
\caption{\small Number of ULXs per galaxy as a function of the host's SFR. Filled red circles are galaxies from this paper, while open black circles are from \cite{Mapelli11}. Errorbars on the SFR are derived assuming a 50\% uncertainty for all measures as in the previous papers. The red dashed line is the power-law fit obtained assuming that the index of the power law is equal to 1.
\label{Fig:NulxSFR} }
\end{figure}

In late-type galaxies, there is a strong and almost linear correlation between the number of ULXs per host galaxy and the SFR of the host galaxy (e.g. \citealt{Swartz09,Mapelli09,Mapelli10,Mapelli11,Mineo11}). This has been interpreted as the smoking gun for the association of ULXs with high-mass X-ray binaries in regions of recent star formation. In Fig.~\ref{Fig:NulxSFR}, we compare the galaxies considered in this paper with the sample of 66 late-type galaxies reported by \cite{Mapelli11}, which is an update of the sample published by \citealt{Mapelli10}. Given the large uncertainty of the SFR we plot an errobar at 50\% of the SFR value, for consistency with \citet{Mapelli10}.  With the addition of the ring galaxies studied in this paper the total correlation remains consistent with that reported by \cite{Mapelli11}. Moreover, our galaxies occupy the high SFR and high ULX number region of the plane covered by the comparison sample.

\begin{figure}[h]
\includegraphics[width=9.cm]{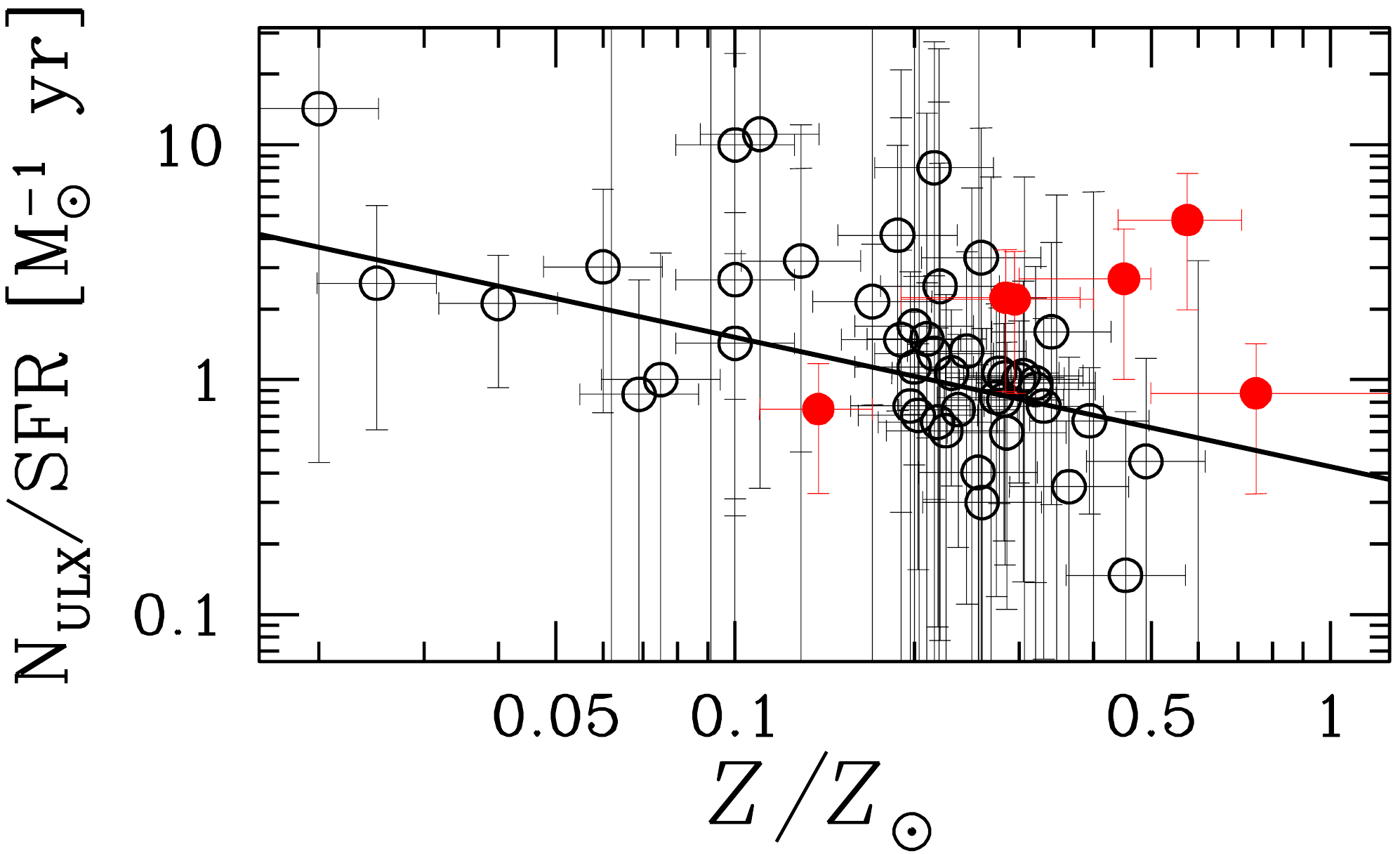}
\caption{\small Number of ULXs per host galaxy, normalized to the SFR, versus host's metalicity ($Z$). Filled red circles are galaxies from this paper, while open black circles are from \cite{Mapelli11}. We assume $Z_\odot=0.02$. The solid black line is the power-law fit from \cite{Mapelli11}. Adding the new galaxies does not change the fit significantly with respect to \cite{Mapelli11} (the power-law index changes from $-0.55^{+0.21}_{-0.19}$ to $-0.50^{+0.22}_{-0.21}$).  The plotted uncertainties in the RiGs values reflect the latitude of the measures. True uncertainties are most probably larger. Arp~148 is not shown because it lacks a metallicity measurement.
\label{Fig:NulxSFRZ} }
\end{figure}

Several studies (\citealt{Mapelli09,Mapelli10,Mapelli11,Kaaret11,Prestwich13,Brorby14,Brorby16}) also suggest that there is an anti-correlation between the number of ULXs per  host galaxy (normalized to the SFR) and the metallicity $Z$ of the host galaxy. The interpretation of this result is still debated. This might indicate that a large fraction of ULXs are associated with massive stellar black holes ($>20$ M$_\odot$), which are expected to form at sub-solar metallicity (e.g. \citealt{Mapelli09,Zampieri09,Mapelli13,Mapelli14}). Alternatively, it might suggest that metal-poor X-ray binaries are more luminous than their metal-rich peers (e.g. \citealt{Linden10}), which could be the case for both BH or NS counterparts.

\startlongtable
\begin{deluxetable*}{lrrrrc}
\tablecaption{List of sources from the literature and used to compute the XLF}
\label{Tab:oldsrc}
\tablehead{
\colhead{Name} & \colhead{logL$_X^{orig}$/L$_X^{orig}$} & \colhead{L$_X$(0.5-10 keV)} & \colhead{Net Counts} & \colhead{XLF corr}& \colhead{Notes}\\
\colhead{} & \colhead{erg s$^{-1}$} & \colhead{$\times 10^{39}$ erg s$^{-1}$} & \colhead{} & \colhead{*} & \colhead{}
}
\startdata
Cart-N6  &   1.41  &  3.36 &19.97$\pm$4.69    &1.0 & (a,b)\\
Cart-N7  &   5.01  & 11.86 &70.64$\pm$8.60    &1.0 & counts in 0.3-7 keV\\
Cart-N9  &   4.68  & 11.15 &66.20$\pm$8.31    &1.0 &\\
Cart-N10 &  27.54  & 65.49 &383.77$\pm$19.77  &1.0 &\\
Cart-N12 &   3.09  &  7.43 &43.75$\pm$6.78    &1.0 &\\
Cart-N13 &   2.69  &  6.37 &38.02$\pm$6.40    &1.0 &\\
Cart-N14 &   7.59  & 17.70 &105.92$\pm$10.54  &1.0 &\\
Cart-N15 &   1.15  &  2.66 &16.06$\pm$4.24    &1.0 &\\
Cart-N16 &   6.76  & 15.93 &94.51$\pm$9.90    &1.0 &\\
Cart-N17 &   8.32  & 19.47 &116.10$\pm$10.95  &1.0 &\\
Cart-N20 &   0.93  &  2.30 &13.20$\pm$3.87    &1.0 &\\
Cart-N21 &   0.50  &  1.19 &6.97$\pm$2.83     &1.25 &\\
Cart-N22 &   1.02  &  2.47 &14.28$\pm$4.00    &1.0 &\\
Cart-N23 &   0.60  &  1.45 &8.54$\pm$3.16     &1.10 &\\
Cart-N24 &   0.37  &  0.86 &5.16$\pm$2.45     &1.68 &\\
\tableline
N922-2	 &   0.68  &    0.69 &   8.92 & 1.19 & (c)\\  
N922-3	 &   3.87  &    3.92 &  46.69 & 1.0 & counts corrected for PSF in 0.3-8 keV\\
N922-4	 &  24.20  &   24.49 & 341.08 & 1.0 & \\
N922-5	 &   1.19  &    1.20 &  14.57 & 1.05 &\\
N922-6	 &   1.02  &    1.03 &  13.08 & 1.10 &\\
N922-7	 &   0.52  &    0.53 &   6.24 & 1.67 &\\
N922-8	 &   3.10  &    3.14 &  44.31 & 1.0 &\\
N922-9	 &   0.98  &    0.99 &  13.08 & 1.10 &\\
N922-11	 &   3.91  &    3.96 &  43.42 & 1.0 &\\        
N922-12	 &   8.67  &    8.77 & 121.33 & 1.0 &\\
N922-13	 &   0.88  &    0.89 &  12.48 & 1.14 &\\
N922-14	 &   0.64  &    0.65 &   7.73 & 1.33 &\\
\tableline
Arp147-Ring1 &   6.2 &    7.52  & 14.7+4.2-0.5  & 1.25&(d)\\
Arp147-Ring2 &   2.7 &	  3.39  &  6.7+3.0-2.3  & 2.5 & counts in 0.5-8 kEV\\
Arp147-Ring3 &   1.4 &	  1.47  &  2.7+2.1-1.4  & 10 &\\
Arp147-Ring4 &   1.6 &	  2.06  &  3.7+2.4-1.7  & 5 &\\
Arp147-Ring5 &   5.1 &	  3.61  &  7.4+3.2-2.5  & 2.2 &\\ 
Arp147-Ring6 &   1.9 &	  2.43  &  4.5+2.6-1.9  & 3.3 &\\
Arp147-Ring7 &   6.9 &	  5.63  & 10.6+3.7-3.0  & 1.7 &\\
Arp147-Ring8 &   6.7 &	  6.25  & 11.7+3.8-3.0  & 1.42 &\\
Arp147-Ring9 &   5.7 &	  5.54  & 10.5+3.7-3.0  & 1.7 &\\
\tableline
Arp284-4  &     64      &     64.1 & 2657$\pm$52  & 1.0 &(e,f) \\      
Arp284-5  &      0.5    &      0.5 &   21$\pm$5   & 1.0 & count in 0.3-8 keV\\
Arp284-6  &      0.6    &      0.6 &   21$\pm$5   & 1.0&\\
Arp284-7  &      0.7    &      0.7 &   26$\pm$5   & 1.0&\\
Arp284-8  &      0.8    &      0.8 &   30$\pm$6   & 1.0&\\
Arp284-9  &      0.6    &      0.6 &   22$\pm$5   & 1.0&\\
Arp284-10 &      0.3    &      0.3 &   10$\pm$3   & 1.05 &\\
Arp284-11 &      0.8    &      0.8 &   30$\pm$6   & 1.0&\\
Arp284-12 &    24	    &     27.3 & 1076$\pm$33  & 1.0&\\
\enddata
\tablenotetext{a}{Originally published as Log L$_X$}
\tablenotetext{b}{N$_{\rm H} = 1.9 \times 10^{21}$ cm$^{-2}$}
\tablenotetext{c}{N$_{\rm H} = 1.6 \times 10^{20}$ cm$^{-2}$}
\tablenotetext{d}{N$_{\rm H} = 6.2 \times 10^{20}$ cm$^{-2}$}
\tablenotetext{e}{Originally published in units of $10^{38}$ erg s$^{-1}$}
\tablenotetext{f}{N$_{\rm H} = 1.8 \times 10^{21}$ cm$^{-2}$}
\tablenotetext{*}{Multiplicative correction as derived from \citet{KimFabbiano03}}
\tablecomments{The table reports the original published L$_X$ and the unabsorbed L$_X$ in the 0.5-10 keV band computed assuming $\Gamma =1.7$ and Galactic N$_{\rm H}$ as in Notes (and Table 1).}
\end{deluxetable*}


\begin{table*}
\begin{center} 
\caption{}{Upper limits in flux for each observation and relative corresponding luminosity at the distance of the individual galaxies.}
\label{Tab:ulim}
\begin{tabular}{lrrrr}
\tableline\tableline
Name  & f$_{\rm X}$ (0.5-2 keV) & L$_{\rm X}$ (0.5-2 keV) & Area$_{\rm ring}$  & Contamination  \\
   & 10$^{-15}$ erg cm$^{-2} s^{-1}$ & 10$^{39}$ erg cm$^{-2}$& $^{\prime\prime}$  &  src/galaxy\\
\tableline
Cartwheel Galaxy & 0.18 & 0.32 & 2478 & 0.43  \\  
NGC 922          & 0.80 & 0.20 & 3888 & 0.27  \\ 
Arp 147          & 2.7 & 0.54&  294 & 0.009  \\  
AM~0644$-$741    & 0.83 & 0.78 & 3010 & 0.20  \\  
Arp 143          & 1.2 & 0.44 & 2032 & 0.11  \\ 
Arp 148          & 1.1 & 2.63 &  282 & 0.016  \\ 
Arp 284          & 0.8 & 0.44 & 1399 & 0.10  \\ 
\tableline
\end{tabular}
\end{center}
\end{table*}

Fig.~\ref{Fig:NulxSFRZ} shows the number of ULXs (normalized to the SFR) versus the metallicity for six of the galaxies in our sample (Arp~148 is not included because it lacks a metallicity measurement), compared to the sample by \cite{Mapelli11}. The overall trend is still the same, and our RiGs seem to sit at the high N$_{ULX}$ and high metallicity end. While this seems a tantalizing result, it is important to stress that most galaxies in our sample (with the exception of the Cartwheel) lack an accurate metallicity estimate, while the sample of \cite{Mapelli11} was selected on the basis of the availability of a reliable metallicity estimate. We stress that the true uncertainties on the metallicity content is probably much higher than what plotted here, which reflects the maximum and minimum measure, or 0.1dex, whichever is larger.

We investigate if the galaxy impact has a direct effect on the nuclear activity of the target galaxies. Most, but not all the ring galaxies show emission in the region corresponding to the original nucleus 
of the galaxy before the impact. In two galaxies it is not straightforward to distinguish 
between the circumstellar diffuse emission of the galaxy and a central nuclear source (NGC 922) 
or it is not obvious where the optical nucleus currently is (Arp 147). The impact, in fact, might have been
slightly different for different galaxies and/or projection effects might make the nucleus appear projected onto the ring \citep{gerber92,Mapelli12}.  
The presence of an active nucleus is not ubiquitous: the Cartwheel Galaxy for instance
does not have one. 
With the small statistics available we infer about a 60\% chance of finding 
an X-ray faint nucleus (L$_X = 10^{39-40}$ erg/s) in RiGs, implying that the effect of the impact on the activity of the nucleus, if present, is mild.

As discussed in Sect.~\ref{xlf}, we find a substantial agreement with previous measures of the XLF (e.g. \citet{Swartz11}), with a hint to a flatter slope and a larger number of bright ULXs in RiGs. An interesting comparison could be made with the results of \cite{Wang16} (similar analyses were done previously, with smaller samples, by \citet{kilgard02,kilgard05}) for more than 300 galaxies spanning all morphological types observed in the Chandra ACIS catalog.  These authors find a decrease in the steepness of the XLF going from early to late types, with peculiar and irregular being the flattest. The high L$_X$ sample is dominated by late type objects, which in turn is dominated by spirals. The peculiar galaxies slope, to which RiGs should better compare, is $\alpha = 0.55 \pm 0.30$, consistent with findings in this paper and in \citet{Swartz11}.

Given the recent discovery of PULXs (see Sect.~\ref{intro}), it is tempting to speculate about the composition of the XLF in this sample. We do not really know how to distinguish observationally PULXs, unless pulsation are seen, which might be a rare event also for known PULXs. A first attempt at proposing a discrimination has been based on the spectral shape of the X-ray emission \citep{Pintore17}. In our case, however, the sources are too faint to measure the spectral shape of single sources. The spectrum of the collective sample of detected point sources has a slope of $\sim 1.6-1.7$ which does not allow us to discriminate. 
A bimodal flux distribution has been observed in a few PULXs, which might depend on the propeller effect \citep{Tsygankov16,Israel17a,Israel17b} or on changes in the accretion rate leading to accretion flow sphericization \citep{Grebenev17}. To establish the presence of a low flux state is a very observationally consuming task.

The collective sample of RiGs seems to confirm the effect of low metallicity on the production of ULXs, especially if compared with a sample of nearby Luminous Infrared Galaxies \citep{Luangtip15} selected to have a large SFR ($> 7$ M$_{\odot}$/yr) which shows a large deficit in ULX per unit SFR, consistent also with the results of the high metallicity sample from \citet{Prestwich13}. The luminous infrared galaxies in the \cite{Luangtip15} sample have in fact metallicity generally $Z > Z_{\odot}$, however the authors claim a significant contribution from dust which obscures the emission of the fainter sources and impedes their detection. The issue is possibly more complex and better and more adequate measures - e.g. metallicity at different position in the galaxy -   are necessary to address it.

\section{Conclusions}

We have collected all the "bright and famous" ring galaxies observed by Chandra with the aim of studying their point sources population detected along the ring. We use both our analysis and literature data to gather an homogeneous sample of 50 ULXs, while 63 sources in total are detected along the seven rings. We construct the first ring-only XLF. This is an estimate of the ULX luminosity distribution at a young age after the burst of star formation, even if uncertainties in the age of the stellar population are still large, both theoretically and observationally. However the young age of the encounter favors both high mass donors and BH accretor, since we should be observing these galaxies at most a few hundreds Myr after the beginning of the star formation episode (see e.g. \cite{Renaud18}). We see a larger average number of ULXs per galaxy than in other galaxies, even if there is not a statistical distinction with the bulk of spiral galaxies XLF. 
We are tempted to favor a high mass BH interpretation for the brightest objects, given their relative higher luminosities, given also the recent discovery of BHs of mass in excess of 60 M$_{\odot}$ due to their gravitational wave emission from coalescence by the LIGO-Virgo collaboration \citep{Abbott16a,Abbott17a,Abbott17b}.
RiGs, the Cartwheel in particular due to its large number of ULXs, are perfect test-bed for models of binary formation and the origin of ULX, such as the one from \cite{Wiktorowicz17}.

\acknowledgments

We thank Alberto Moretti, Doug Swartz, Grzegorz Wiktorowicz and Andreas Zezas for useful discussions and inputs. We also thank the referee for useful comments that helped improve the paper. 
This research has made use of data obtained from the Chandra Data Archive and software provided by the Chandra X-ray Center (CXC) in the application packages DS9, CIAO, ChIPS, and Sherpa. The CXC is operated for NASA by the Smithsonian Astrophysical Observatory.
Optical images were taken with the NASA/ESA Hubble Space Telescope, and obtained from the Hubble Heritage program and Hubble Legacy Archive, which is a collaboration between the Space Telescope Science Institute (STScI/NASA), the Space Telescope European Coordinating Facility (ST-ECF/ESA) and the Canadian Astronomy Data Centre (CADC/NRC/CSA). 
Funding for SDSS-III (we acknowledge here use of one image) has been provided by the Alfred P. Sloan Foundation, the Participating Institutions, the National Science Foundation, and the U.S. Department of Energy Office of Science. 
This research has made use of the NASA/IPAC Extragalactic Database (NED),
which is operated by the Jet Propulsion Laboratory, Caltech,under contract with NASA.
AW acknowledges financial support from ASI through the ASI-INAF agreements 2015-023-R.0 and 2017-14-H.0.
AF acknowledges support by the CXC under NASA contract NAS8-03060.
MM acknowledges financial support from INAF through PRIN-SKA  "Opening a new era in pulsars and compact objects science with MeerKat" and from the MERAC Foundation through the project "The physics of gas and protoplanetary discs in the Galactic centre".

\software{CIAO \citep{fruscione06}, Sherpa \citep{Freeman01,Doe07}, 
ChIPS, \citep{Miller15}, DS9 \citep{Joye11}, XSPEC \citep{Arnaud96}, smongo \textsc{\ https://www.astro.princeton.edu/$\sim$rhl/sm/}}


\bibliographystyle{aasjournal}
\bibliography{bibliography}

\begin{thebibliography}{}
\expandafter\ifx\csname natexlab\endcsname\relax\def\natexlab#1{#1}\fi

\bibitem[{{Abbott} {et~al.}(2016){Abbott}, {Abbott}, {Abbott}, {Abernathy},
  {Acernese}, {Ackley}, {Adams}, {Adams}, {Addesso}, {Adhikari}, \&
  et~al.}]{Abbott16a}
{Abbott}, B.~P., {Abbott}, R., {Abbott}, T.~D., {et~al.} 2016, \prd, 93, 122003

\bibitem[{{Abbott} {et~al.}(2017{\natexlab{a}}){Abbott}, {Abbott}, {Abbott},
  {Acernese}, {Ackley}, {Adams}, {Adams}, {Addesso}, {Adhikari}, {Adya}, \&
  et~al.}]{Abbott17a}
---. 2017{\natexlab{a}}, Physical Review Letters, 118, 221101

\bibitem[{{Abbott} {et~al.}(2017{\natexlab{b}}){Abbott}, {Abbott}, {Abbott},
  {Acernese}, {Ackley}, {Adams}, {Adams}, {Addesso}, {Adhikari}, {Adya}, \&
  et~al.}]{Abbott17b}
---. 2017{\natexlab{b}}, Physical Review Letters, 119, 141101

\bibitem[{{Abbott} {et~al.}(2017{\natexlab{c}}){Abbott}, {Abbott}, {Abbott},
  {Acernese}, {Ackley}, {Adams}, {Adams}, {Addesso}, {Adhikari}, {Adya}, \&
  et~al.}]{Abbott17c}
---. 2017{\natexlab{c}}, \prd, 96, 022001

\bibitem[{{Abolfathi} {et~al.}(2017){Abolfathi}, {Aguado}, {Aguilar}, {Allende
  Prieto}, {Almeida}, {Tasnim Ananna}, {Anders}, {Anderson}, {Andrews},
  {Anguiano}, \& et~al.}]{Abolfathi17}
{Abolfathi}, B., {Aguado}, D.~S., {Aguilar}, G., {et~al.} 2017, ArXiv e-prints,
  arXiv:1707.09322

\bibitem[{{Anders} \& {Grevesse}(1989)}]{anders1989}
{Anders}, E., \& {Grevesse}, N. 1989, \gca, 53, 197

\bibitem[{{Appleton} {et~al.}(1987){Appleton}, {Ghigo}, {van Gorkom},
  {Schombert}, \& {Struck-Marcell}}]{Appleton87}
{Appleton}, P.~N., {Ghigo}, F.~D., {van Gorkom}, J.~H., {Schombert}, J.~M., \&
  {Struck-Marcell}, C. 1987, \nat, 330, 140

\bibitem[{{Appleton} \& {Marston}(1997)}]{Appleton97}
{Appleton}, P.~N., \& {Marston}, A.~P. 1997, \aj, 113, 201

\bibitem[{{Arnaud}(1996)}]{Arnaud96}
{Arnaud}, K.~A. 1996, in Astronomical Society of the Pacific Conference Series,
  Vol. 101, Astronomical Data Analysis Software and Systems V, ed. G.~H.
  {Jacoby} \& J.~{Barnes}, 17

\bibitem[{{Athanassoula} \& {Bosma}(1985)}]{Athanassoula85}
{Athanassoula}, E., \& {Bosma}, A. 1985, \araa, 23, 147

\bibitem[{{Bachetti} {et~al.}(2014){Bachetti}, {Harrison}, {Walton},
  {Grefenstette}, {Chakrabarty}, {F{\"u}rst}, {Barret}, {Beloborodov}, {Boggs},
  {Christensen}, {Craig}, {Fabian}, {Hailey}, {Hornschemeier}, {Kaspi},
  {Kulkarni}, {Maccarone}, {Miller}, {Rana}, {Stern}, {Tendulkar}, {Tomsick},
  {Webb}, \& {Zhang}}]{Bachetti14}
{Bachetti}, M., {Harrison}, F.~A., {Walton}, D.~J., {et~al.} 2014, \nat, 514,
  202

\bibitem[{{Beir{\~a}o} {et~al.}(2009){Beir{\~a}o}, {Appleton}, {Brandl},
  {Seibert}, {Jarrett}, \& {Houck}}]{beirao09}
{Beir{\~a}o}, P., {Appleton}, P.~N., {Brandl}, B.~R., {et~al.} 2009, \apj, 693,
  1650

\bibitem[{{Bogd{\'a}n} \& {Gilfanov}(2011)}]{bogdan11}
{Bogd{\'a}n}, {\'A}., \& {Gilfanov}, M. 2011, \mnras, 418, 1901

\bibitem[{{Brorby} {et~al.}(2014){Brorby}, {Kaaret}, \& {Prestwich}}]{Brorby14}
{Brorby}, M., {Kaaret}, P., \& {Prestwich}, A. 2014, \mnras, 441, 2346

\bibitem[{{Brorby} {et~al.}(2016){Brorby}, {Kaaret}, {Prestwich}, \&
  {Mirabel}}]{Brorby16}
{Brorby}, M., {Kaaret}, P., {Prestwich}, A., \& {Mirabel}, I.~F. 2016, \mnras,
  457, 4081

\bibitem[{{Carpano} {et~al.}(2018){Carpano}, {Haberl}, {Maitra}, \&
  {Vasilopoulos}}]{Carpano18}
{Carpano}, S., {Haberl}, F., {Maitra}, C., \& {Vasilopoulos}, G. 2018, \mnras,
  476, L45

\bibitem[{{Crivellari} {et~al.}(2009){Crivellari}, {Wolter}, \&
  {Trinchieri}}]{Crivellari09}
{Crivellari}, E., {Wolter}, A., \& {Trinchieri}, G. 2009, \aap, 501, 445

\bibitem[{{Doe} {et~al.}(2007){Doe}, {Nguyen}, {Stawarz}, {Refsdal},
  {Siemiginowska}, {Burke}, {Evans}, {Evans}, {McDowell}, {Houck}, \&
  {Nowak}}]{Doe07}
{Doe}, S., {Nguyen}, D., {Stawarz}, C., {et~al.} 2007, in Astronomical Society
  of the Pacific Conference Series, Vol. 376, Astronomical Data Analysis
  Software and Systems XVI, ed. R.~A. {Shaw}, F.~{Hill}, \& D.~J. {Bell}, 543

\bibitem[{{Farrell} {et~al.}(2009){Farrell}, {Webb}, {Barret}, {Godet}, \&
  {Rodrigues}}]{Farrell09}
{Farrell}, S.~A., {Webb}, N.~A., {Barret}, D., {Godet}, O., \& {Rodrigues},
  J.~M. 2009, \nat, 460, 73

\bibitem[{{Feng} \& {Kaaret}(2008)}]{Feng08}
{Feng}, H., \& {Kaaret}, P. 2008, \apj, 675, 1067

\bibitem[{{Feng} {et~al.}(2010){Feng}, {Rao}, \& {Kaaret}}]{feng10}
{Feng}, H., {Rao}, F., \& {Kaaret}, P. 2010, \apjl, 710, L137

\bibitem[{{Feng} \& {Soria}(2011)}]{Feng11}
{Feng}, H., \& {Soria}, R. 2011, \nar, 55, 166

\bibitem[{{Few} {et~al.}(1982){Few}, {Arp}, \& {Madore}}]{Few82}
{Few}, J.~M.~A., {Arp}, H.~C., \& {Madore}, B.~F. 1982, \mnras, 199, 633

\bibitem[{{Fogarty} {et~al.}(2011){Fogarty}, {Thatte}, {Tecza}, {Clarke},
  {Goodsall}, {Houghton}, {Salter}, {Davies}, \& {Kassin}}]{Fogarty11}
{Fogarty}, L., {Thatte}, N., {Tecza}, M., {et~al.} 2011, \mnras, 417, 835

\bibitem[{{Fosbury} \& {Hawarden}(1977)}]{FosburyHawarden77}
{Fosbury}, R.~A.~E., \& {Hawarden}, T.~G. 1977, \mnras, 178, 473

\bibitem[{{Fragos} {et~al.}(2013){Fragos}, {Lehmer}, {Naoz}, {Zezas}, \&
  {Basu-Zych}}]{Fragos13}
{Fragos}, T., {Lehmer}, B.~D., {Naoz}, S., {Zezas}, A., \& {Basu-Zych}, A.
  2013, \apjl, 776, L31

\bibitem[{{Freeman} {et~al.}(2001){Freeman}, {Doe}, \&
  {Siemiginowska}}]{Freeman01}
{Freeman}, P., {Doe}, S., \& {Siemiginowska}, A. 2001, in \procspie, Vol. 4477,
  Astronomical Data Analysis, ed. J.-L. {Starck} \& F.~D. {Murtagh}, 76--87

\bibitem[{{Fruscione} {et~al.}(2006){Fruscione}, {McDowell}, {Allen},
  {Brickhouse}, {Burke}, {Davis}, {Durham}, {Elvis}, {Galle}, {Harris},
  {Huenemoerder}, {Houck}, {Ishibashi}, {Karovska}, {Nicastro}, {Noble},
  {Nowak}, {Primini}, {Siemiginowska}, {Smith}, \& {Wise}}]{fruscione06}
{Fruscione}, A., {McDowell}, J.~C., {Allen}, G.~E., {et~al.} 2006, in
  \procspie, Vol. 6270, Society of Photo-Optical Instrumentation Engineers
  (SPIE) Conference Series, 62701V

\bibitem[{{F{\"u}rst} {et~al.}(2016){F{\"u}rst}, {Walton}, {Harrison}, {Stern},
  {Barret}, {Brightman}, {Fabian}, {Grefenstette}, {Madsen}, {Middleton},
  {Miller}, {Pottschmidt}, {Ptak}, {Rana}, \& {Webb}}]{Fuerst16}
{F{\"u}rst}, F., {Walton}, D.~J., {Harrison}, F.~A., {et~al.} 2016, \apjl, 831,
  L14

\bibitem[{{Gao} {et~al.}(2003){Gao}, {Wang}, {Appleton}, \& {Lucas}}]{Gao03}
{Gao}, Y., {Wang}, Q.~D., {Appleton}, P.~N., \& {Lucas}, R.~A. 2003, \apjl,
  596, L171

\bibitem[{{Garc{\'{\i}}a-Vargas} {et~al.}(1997){Garc{\'{\i}}a-Vargas},
  {Gonz{\'a}lez-Delgado}, {P{\'e}rez}, {Alloin}, {D{\'{\i}}az}, \&
  {Terlevich}}]{Garcia-Vargas97}
{Garc{\'{\i}}a-Vargas}, M.~L., {Gonz{\'a}lez-Delgado}, R.~M., {P{\'e}rez}, E.,
  {et~al.} 1997, \apj, 478, 112

\bibitem[{{Garmire} {et~al.}(2003){Garmire}, {Bautz}, {Ford}, {Nousek}, \&
  {Ricker}}]{garmire03}
{Garmire}, G.~P., {Bautz}, M.~W., {Ford}, P.~G., {Nousek}, J.~A., \& {Ricker},
  Jr., G.~R. 2003, in \procspie, Vol. 4851, X-Ray and Gamma-Ray Telescopes and
  Instruments for Astronomy., ed. J.~E. {Truemper} \& H.~D. {Tananbaum}, 28--44

\bibitem[{{Gerber} {et~al.}(1992){Gerber}, {Lamb}, \& {Balsara}}]{gerber92}
{Gerber}, R.~A., {Lamb}, S.~A., \& {Balsara}, D.~S. 1992, \apjl, 399, L51

\bibitem[{{Gerber} {et~al.}(1996){Gerber}, {Lamb}, \& {Balsara}}]{Gerber96}
---. 1996, \mnras, 278, 345

\bibitem[{{Grebenev}(2017)}]{Grebenev17}
{Grebenev}, S.~A. 2017, Astronomy Letters, 43, 464

\bibitem[{{Grimm} {et~al.}(2003){Grimm}, {Gilfanov}, \& {Sunyaev}}]{Grimm03}
{Grimm}, H.-J., {Gilfanov}, M., \& {Sunyaev}, R. 2003, \mnras, 339, 793

\bibitem[{{Heida} {et~al.}(2013){Heida}, {Jonker}, {Torres}, {Roberts},
  {Miniutti}, {Fabian}, \& {Ratti}}]{Heida13}
{Heida}, M., {Jonker}, P.~G., {Torres}, M.~A.~P., {et~al.} 2013, \mnras, 433,
  681

\bibitem[{{Higdon}(1995)}]{Higdon95}
{Higdon}, J.~L. 1995, \apj, 455, 524

\bibitem[{{Higdon}(1996)}]{Higdon96}
---. 1996, \apj, 467, 241

\bibitem[{{Higdon} {et~al.}(2015){Higdon}, {Higdon}, {Mart{\'{\i}}n Ruiz}, \&
  {Rand}}]{Higdon15}
{Higdon}, J.~L., {Higdon}, S.~J.~U., {Mart{\'{\i}}n Ruiz}, S., \& {Rand}, R.~J.
  2015, \apjl, 814, L1

\bibitem[{{Higdon} {et~al.}(2011){Higdon}, {Higdon}, \& {Rand}}]{Higdon11}
{Higdon}, J.~L., {Higdon}, S.~J.~U., \& {Rand}, R.~J. 2011, \apj, 739, 97

\bibitem[{{Higdon} {et~al.}(1997){Higdon}, {Rand}, \& {Lord}}]{Higdon97}
{Higdon}, J.~L., {Rand}, R.~J., \& {Lord}, S.~D. 1997, \apjl, 489, L133

\bibitem[{{Higdon} \& {Wallin}(1997)}]{HigdonWallin97}
{Higdon}, J.~L., \& {Wallin}, J.~F. 1997, \apj, 474, 686

\bibitem[{{Israel} {et~al.}(2017{\natexlab{a}}){Israel}, {Belfiore}, {Stella},
  {Esposito}, {Casella}, {De Luca}, {Marelli}, {Papitto}, {Perri}, {Puccetti},
  {Castillo}, {Salvetti}, {Tiengo}, {Zampieri}, {D'Agostino}, {Greiner},
  {Haberl}, {Novara}, {Salvaterra}, {Turolla}, {Watson}, {Wilms}, \&
  {Wolter}}]{Israel17b}
{Israel}, G.~L., {Belfiore}, A., {Stella}, L., {et~al.} 2017{\natexlab{a}},
  Science, 355, 817

\bibitem[{{Israel} {et~al.}(2017{\natexlab{b}}){Israel}, {Papitto}, {Esposito},
  {Stella}, {Zampieri}, {Belfiore}, {Rodr{\'{\i}}guez Castillo}, {De Luca},
  {Tiengo}, {Haberl}, {Greiner}, {Salvaterra}, {Sandrelli}, \&
  {Lisini}}]{Israel17a}
{Israel}, G.~L., {Papitto}, A., {Esposito}, P., {et~al.} 2017{\natexlab{b}},
  \mnras, 466, L48

\bibitem[{{Jeske}(1986)}]{Jeske86}
{Jeske}, N.~A. 1986, PhD thesis, California Univ., Berkeley.

\bibitem[{{Joye}(2011)}]{Joye11}
{Joye}, W. 2011, in Astronomical Data Analysis Software and Systems XX. ASP
  Conference Proceedings, Vol. 442, proceedings of a Conference held at Seaport
  World Trade Center, Boston, Massachusetts, USA on 7-11 November 2010. Edited
  by Ian N. Evans, Alberto Accomazzi, Douglas J. Mink, and Arnold H. Rots. San
  Francisco: Astronomical Society of the Pacific, 2011., p.633, Vol. 442, 633

\bibitem[{{Kaaret} {et~al.}(2017){Kaaret}, {Feng}, \& {Roberts}}]{kaaret17}
{Kaaret}, P., {Feng}, H., \& {Roberts}, T.~P. 2017, \araa, 55, 303

\bibitem[{{Kaaret} {et~al.}(2011){Kaaret}, {Schmitt}, \& {Gorski}}]{Kaaret11}
{Kaaret}, P., {Schmitt}, J., \& {Gorski}, M. 2011, \apj, 741, 10

\bibitem[{{Kilgard} {et~al.}(2002){Kilgard}, {Kaaret}, {Krauss}, {Prestwich},
  {Raley}, \& {Zezas}}]{kilgard02}
{Kilgard}, R.~E., {Kaaret}, P., {Krauss}, M.~I., {et~al.} 2002, \apj, 573, 138

\bibitem[{{Kilgard} {et~al.}(2005){Kilgard}, {Cowan}, {Garcia}, {Kaaret},
  {Krauss}, {McDowell}, {Prestwich}, {Primini}, {Stockdale}, {Trinchieri},
  {Ward}, \& {Zezas}}]{kilgard05}
{Kilgard}, R.~E., {Cowan}, J.~J., {Garcia}, M.~R., {et~al.} 2005, \apjs, 159,
  214

\bibitem[{{Kim} \& {Fabbiano}(2003)}]{KimFabbiano03}
{Kim}, D.-W., \& {Fabbiano}, G. 2003, \apj, 586, 826

\bibitem[{{King} \& {Lasota}(2016)}]{King16}
{King}, A., \& {Lasota}, J.-P. 2016, \mnras, 458, L10

\bibitem[{{Lan{\c c}on} {et~al.}(2001){Lan{\c c}on}, {Goldader}, {Leitherer},
  \& {Gonz{\'a}lez Delgado}}]{Lancon01}
{Lan{\c c}on}, A., {Goldader}, J.~D., {Leitherer}, C., \& {Gonz{\'a}lez
  Delgado}, R.~M. 2001, \apj, 552, 150

\bibitem[{{Linden} {et~al.}(2010){Linden}, {Kalogera}, {Sepinsky}, {Prestwich},
  {Zezas}, \& {Gallagher}}]{Linden10}
{Linden}, T., {Kalogera}, V., {Sepinsky}, J.~F., {et~al.} 2010, \apj, 725, 1984

\bibitem[{{Liu} {et~al.}(2013){Liu}, {Bregman}, {Bai}, {Justham}, \&
  {Crowther}}]{Liu13}
{Liu}, J.-F., {Bregman}, J.~N., {Bai}, Y., {Justham}, S., \& {Crowther}, P.
  2013, \nat, 503, 500

\bibitem[{{Luangtip} {et~al.}(2015){Luangtip}, {Roberts}, {Mineo}, {Lehmer},
  {Alexander}, {Jackson}, {Goulding}, \& {Fischer}}]{Luangtip15}
{Luangtip}, W., {Roberts}, T.~P., {Mineo}, S., {et~al.} 2015, \mnras, 446, 470

\bibitem[{{Lynds} \& {Toomre}(1976)}]{Lynds76}
{Lynds}, R., \& {Toomre}, A. 1976, \apj, 209, 382

\bibitem[{{Mapelli}(2016)}]{Mapelli16}
{Mapelli}, M. 2016, \mnras, 459, 3432

\bibitem[{{Mapelli} {et~al.}(2009){Mapelli}, {Colpi}, \&
  {Zampieri}}]{Mapelli09}
{Mapelli}, M., {Colpi}, M., \& {Zampieri}, L. 2009, \mnras, 395, L71

\bibitem[{{Mapelli} \& {Mayer}(2012)}]{Mapelli12}
{Mapelli}, M., \& {Mayer}, L. 2012, \mnras, 420, 1158

\bibitem[{{Mapelli} {et~al.}(2011){Mapelli}, {Ripamonti}, {Zampieri}, \&
  {Colpi}}]{Mapelli11}
{Mapelli}, M., {Ripamonti}, E., {Zampieri}, L., \& {Colpi}, M. 2011,
  Astronomische Nachrichten, 332, 414

\bibitem[{{Mapelli} {et~al.}(2010){Mapelli}, {Ripamonti}, {Zampieri}, {Colpi},
  \& {Bressan}}]{Mapelli10}
{Mapelli}, M., {Ripamonti}, E., {Zampieri}, L., {Colpi}, M., \& {Bressan}, A.
  2010, \mnras, 408, 234

\bibitem[{{Mapelli} \& {Zampieri}(2014)}]{Mapelli14}
{Mapelli}, M., \& {Zampieri}, L. 2014, \apj, 794, 7

\bibitem[{{Mapelli} {et~al.}(2013){Mapelli}, {Zampieri}, {Ripamonti}, \&
  {Bressan}}]{Mapelli13}
{Mapelli}, M., {Zampieri}, L., {Ripamonti}, E., \& {Bressan}, A. 2013, \mnras,
  429, 2298

\bibitem[{{Mayya} {et~al.}(2005){Mayya}, {Bizyaev}, {Romano}, {Garcia-Barreto},
  \& {Vorobyov}}]{Mayya05}
{Mayya}, Y.~D., {Bizyaev}, D., {Romano}, R., {Garcia-Barreto}, J.~A., \&
  {Vorobyov}, E.~I. 2005, \apjl, 620, L35

\bibitem[{{Mesinger} {et~al.}(2013){Mesinger}, {Ferrara}, \&
  {Spiegel}}]{Mesinger13}
{Mesinger}, A., {Ferrara}, A., \& {Spiegel}, D.~S. 2013, \mnras, 431, 621

\bibitem[{{Mezcua} {et~al.}(2015){Mezcua}, {Roberts}, {Lobanov}, \&
  {Sutton}}]{mezcua15}
{Mezcua}, M., {Roberts}, T.~P., {Lobanov}, A.~P., \& {Sutton}, A.~D. 2015,
  \mnras, 448, 1893

\bibitem[{{Miller} {et~al.}(2015){Miller}, {Burke}, {Evans}, {Evans}, \&
  {McLaughlin}}]{Miller15}
{Miller}, J., {Burke}, D.~J., {Evans}, I.~N., {Evans}, J.~D., \& {McLaughlin},
  W. 2015, in Astronomical Data Analysis Software an Systems XXIV (ADASS XXIV),
  Proceedings of a conference held 5-9 October 2014 at Calgary, Alberta Canada.
  Edited by A. R. Taylor and E. Rosolowsky. San Francisco: Astronomical Society
  of the Pacific, 2015., p.111, Vol. 495, 111

\bibitem[{{Mineo} {et~al.}(2011){Mineo}, {Gilfanov}, \& {Sunyaev}}]{Mineo11}
{Mineo}, S., {Gilfanov}, M., \& {Sunyaev}, R. 2011, Astronomische Nachrichten,
  332, 349

\bibitem[{{Mineo} {et~al.}(2012){Mineo}, {Gilfanov}, \& {Sunyaev}}]{Mineo12}
---. 2012, \mnras, 419, 2095

\bibitem[{{Moretti} {et~al.}(2003){Moretti}, {Campana}, {Lazzati}, \&
  {Tagliaferri}}]{Moretti03}
{Moretti}, A., {Campana}, S., {Lazzati}, D., \& {Tagliaferri}, G. 2003, \apj,
  588, 696

\bibitem[{{Motch} {et~al.}(2014){Motch}, {Pakull}, {Soria}, {Gris{\'e}}, \&
  {Pietrzy{\'n}ski}}]{Motch14}
{Motch}, C., {Pakull}, M.~W., {Soria}, R., {Gris{\'e}}, F., \&
  {Pietrzy{\'n}ski}, G. 2014, \nat, 514, 198

\bibitem[{{Pellerin} {et~al.}(2010){Pellerin}, {Meurer}, {Bekki}, {Elmegreen},
  {Wong}, \& {Knezek}}]{Pellerin10}
{Pellerin}, A., {Meurer}, G.~R., {Bekki}, K., {et~al.} 2010, \aj, 139, 1369

\bibitem[{{Pintore} {et~al.}(2017){Pintore}, {Zampieri}, {Stella}, {Wolter},
  {Mereghetti}, \& {Israel}}]{Pintore17}
{Pintore}, F., {Zampieri}, L., {Stella}, L., {et~al.} 2017, \apj, 836, 113

\bibitem[{{Pizzolato} {et~al.}(2010){Pizzolato}, {Wolter}, \&
  {Trinchieri}}]{Pizzolato10}
{Pizzolato}, F., {Wolter}, A., \& {Trinchieri}, G. 2010, \mnras, 406, 1116

\bibitem[{{Portegies Zwart} {et~al.}(2004){Portegies Zwart}, {Dewi}, \&
  {Maccarone}}]{portegieszwart04}
{Portegies Zwart}, S.~F., {Dewi}, J., \& {Maccarone}, T. 2004, \mnras, 355, 413

\bibitem[{{Prestwich} {et~al.}(2013){Prestwich}, {Tsantaki}, {Zezas},
  {Jackson}, {Roberts}, {Foltz}, {Linden}, \& {Kalogera}}]{Prestwich13}
{Prestwich}, A.~H., {Tsantaki}, M., {Zezas}, A., {et~al.} 2013, \apj, 769, 92

\bibitem[{{Prestwich} {et~al.}(2012){Prestwich}, {Galache}, {Linden},
  {Kalogera}, {Zezas}, {Roberts}, {Kilgard}, {Wolter}, \&
  {Trinchieri}}]{Prestwich12}
{Prestwich}, A.~H., {Galache}, J.~L., {Linden}, T., {et~al.} 2012, \apj, 747,
  150

\bibitem[{{Ranalli} {et~al.}(2003){Ranalli}, {Comastri}, \&
  {Setti}}]{Ranalli03}
{Ranalli}, P., {Comastri}, A., \& {Setti}, G. 2003, \aap, 399, 39

\bibitem[{{Rappaport} {et~al.}(2010){Rappaport}, {Levine}, {Pooley}, \&
  {Steinhorn}}]{Rappaport10}
{Rappaport}, S., {Levine}, A., {Pooley}, D., \& {Steinhorn}, B. 2010, \apj,
  721, 1348

\bibitem[{{Renaud} {et~al.}(2018){Renaud}, {Athanassoula}, {Amram}, {Bosma},
  {Bournaud}, {Duc}, {Epinat}, {Fensch}, {Kraljic}, {Perret}, \&
  {Struck}}]{Renaud18}
{Renaud}, F., {Athanassoula}, E., {Amram}, P., {et~al.} 2018, \mnras, 473, 585

\bibitem[{{Romano} {et~al.}(2008){Romano}, {Mayya}, \& {Vorobyov}}]{Romano08}
{Romano}, R., {Mayya}, Y.~D., \& {Vorobyov}, E.~I. 2008, \aj, 136, 1259

\bibitem[{{Schmitt} {et~al.}(2006){Schmitt}, {Calzetti}, {Armus}, {Giavalisco},
  {Heckman}, {Kennicutt}, {Leitherer}, \& {Meurer}}]{Schmitt06}
{Schmitt}, H.~R., {Calzetti}, D., {Armus}, L., {et~al.} 2006, \apjs, 164, 52

\bibitem[{{Smith} {et~al.}(2005){Smith}, {Struck}, \& {Nowak}}]{Smith05}
{Smith}, B.~J., {Struck}, C., \& {Nowak}, M.~A. 2005, \aj, 129, 1350

\bibitem[{{Strauss} {et~al.}(1992){Strauss}, {Huchra}, {Davis}, {Yahil},
  {Fisher}, \& {Tonry}}]{Strauss92}
{Strauss}, M.~A., {Huchra}, J.~P., {Davis}, M., {et~al.} 1992, \apjs, 83, 29

\bibitem[{{Swartz} {et~al.}(2011){Swartz}, {Soria}, {Tennant}, \&
  {Yukita}}]{Swartz11}
{Swartz}, D.~A., {Soria}, R., {Tennant}, A.~F., \& {Yukita}, M. 2011, \apj,
  741, 49

\bibitem[{{Swartz} {et~al.}(2009){Swartz}, {Tennant}, \& {Soria}}]{Swartz09}
{Swartz}, D.~A., {Tennant}, A.~F., \& {Soria}, R. 2009, \apj, 703, 159

\bibitem[{{Tsygankov} {et~al.}(2016){Tsygankov}, {Mushtukov}, {Suleimanov}, \&
  {Poutanen}}]{Tsygankov16}
{Tsygankov}, S.~S., {Mushtukov}, A.~A., {Suleimanov}, V.~F., \& {Poutanen}, J.
  2016, \mnras, 457, 1101

\bibitem[{{van der Marel}(2004)}]{vanderMarel04}
{van der Marel}, R.~P. 2004, Coevolution of Black Holes and Galaxies, 37

\bibitem[{{Walton} {et~al.}(2018){Walton}, {F{\"u}rst}, {Harrison}, {Stern},
  {Bachetti}, {Barret}, {Brightman}, {Fabian}, {Middleton}, {Ptak}, \&
  {Tao}}]{Walton18}
{Walton}, D.~J., {F{\"u}rst}, F., {Harrison}, F.~A., {et~al.} 2018, \mnras,
  473, 4360

\bibitem[{{Wang} {et~al.}(2016){Wang}, {Qiu}, {Liu}, \& {Bregman}}]{Wang16}
{Wang}, S., {Qiu}, Y., {Liu}, J., \& {Bregman}, J.~N. 2016, \apj, 829, 20

\bibitem[{{Wiktorowicz} {et~al.}(2017){Wiktorowicz}, {Sobolewska}, {Lasota}, \&
  {Belczynski}}]{Wiktorowicz17}
{Wiktorowicz}, G., {Sobolewska}, M., {Lasota}, J.-P., \& {Belczynski}, K. 2017,
  \apj, 846, 17

\bibitem[{{Wolter} {et~al.}(2015){Wolter}, {Esposito}, {Mapelli}, {Pizzolato},
  \& {Ripamonti}}]{Wolter15}
{Wolter}, A., {Esposito}, P., {Mapelli}, M., {Pizzolato}, F., \& {Ripamonti},
  E. 2015, \mnras, 448, 781

\bibitem[{{Wolter} \& {Trinchieri}(2004)}]{WT04}
{Wolter}, A., \& {Trinchieri}, G. 2004, \aap, 426, 787

\bibitem[{{Wolter} {et~al.}(2006){Wolter}, {Trinchieri}, \& {Colpi}}]{Wolter06}
{Wolter}, A., {Trinchieri}, G., \& {Colpi}, M. 2006, \mnras, 373, 1627

\bibitem[{{Wolter} {et~al.}(1999){Wolter}, {Trinchieri}, \&
  {Iovino}}]{Wolter99}
{Wolter}, A., {Trinchieri}, G., \& {Iovino}, A. 1999, \aap, 342, 41

\bibitem[{{Wong} {et~al.}(2006){Wong}, {Meurer}, {Bekki}, {Hanish},
  {Kennicutt}, {Bland-Hawthorn}, {Ryan-Weber}, {Koribalski}, {Kilborn},
  {Putman}, {Heiner}, {Webster}, {Allen}, {Dopita}, {Doyle}, {Drinkwater},
  {Ferguson}, {Freeman}, {Heckman}, {Hoopes}, {Knezek}, {Meyer}, {Oey},
  {Seibert}, {Smith}, {Staveley-Smith}, {Thilker}, {Werk}, \& {Zwaan}}]{Wong06}
{Wong}, O.~I., {Meurer}, G.~R., {Bekki}, K., {et~al.} 2006, \mnras, 370, 1607

\bibitem[{{Yue} {et~al.}(2013){Yue}, {Ferrara}, {Salvaterra}, \&
  {Chen}}]{Yue13}
{Yue}, B., {Ferrara}, A., {Salvaterra}, R., \& {Chen}, X. 2013, \mnras, 431,
  383

\bibitem[{{Zampieri} \& {Roberts}(2009)}]{Zampieri09}
{Zampieri}, L., \& {Roberts}, T.~P. 2009, \mnras, 400, 677

\bibitem[{{Zezas} {et~al.}(2007){Zezas}, {Fabbiano}, {Baldi}, {Schweizer},
  {King}, {Rots}, \& {Ponman}}]{Zezas07}
{Zezas}, A., {Fabbiano}, G., {Baldi}, A., {et~al.} 2007, \apj, 661, 135

\end{thebibliography}
\end{document}